\documentclass[final]{ws-ijbc}

\usepackage{graphicx,amsfonts,amsmath,amssymb,MnSymbol}
\usepackage[draft]{changes}

\usepackage[english]{babel}

\usepackage{ucs}
\usepackage[utf8x]{inputenc}

\newcommand{\mb}[1]{\mbox{\bfseries \itshape #1}}
\newcommand{\0}{\Bar{0}}
\newcommand{\1}{\Bar{1}}

\newtheorem{theor}{Theorem}
\newtheorem{defi}{Definition}
\newtheorem{proper}{Property}
\newtheorem{coro}{Corollary}

\begin{document}

\catchline{}{}{}{}{} 

\markboth{Letellier {\it et al.}}{Dynamical taxonomy}

\title{Dynamical Taxonomy: some taxonomic ranks to 
systematically classify every chaotic attractor}

\author{Christophe Letellier}

\address{
Rouen Normandie University --- CORIA, Campus Universitaire du Madrillet,
F-76800 Saint-Etienne du Rouvray, France \\
christophe.letellier@coria.fr}

\author{Nataliya Stankevich}

\address{Laboratory of Topological Methods in Dynamics, National Research
University Higher School of Economics, Bolshaya Pecherskaya str., 25/12,
Nizhny Novgorod 603155, Russia \\
stankevichnv@mail.ru}

\author{Otto E. R\"ossler}

\address{Division of Theoretical Chemistry, University of T\"ubingen,
D-72076 T\"ubingen, Germany}

\maketitle

\begin{history}
\received{\today (accepted for publication on September 12, 2021)}
\end{history}

\begin{abstract}
Characterizing accurately chaotic behaviors is not a trivial problem and must
allow to determine the properties that two given chaotic invariant 
sets share or not. The underlying problem is 
the classification of chaotic regimes, and their labelling. Addressing these
problems corresponds to the development of a dynamical taxonomy, exhibiting the
key properties discriminating the variety of chaotic behaviors discussed in the
abundant literature. Starting from the hierarchy of chaos initially proposed
by one of us, we systematized the description of chaotic regimes observed in 
three- and four-dimensional spaces, which cover a large variety of known (and 
less known) examples of chaos. Starting with the spectrum of Lyapunov 
exponents as the first taxonomic ranks, we extended the description to higher
ranks with some concepts inherited from topology (bounding torus, surface of 
section, first-return map, ...).

By treating extensively the R\"ossler and the Lorenz attractors, we extended
the description of branched manifold --- the highest known taxonomic rank for 
classifying chaotic attractor --- by a linking matrix (or linker) to 
multi-component attractors (bounded by a torus whose genus $g \geq 3$).
\end{abstract}


\section{Introduction}

There are various types of deterministic dynamical behavior which can be
observed. Their nature strongly depends on the dimension of the state space
involved. R\"ossler upended his early program of research in chaos with an
attempt at constructing a hierarchy in providing various examples \cite{Ros83}.
On the occasion of his 50th birthday, Michael Klein and Gerold Baier proposed a
second version, a classification mainly based on the spectrum of Lyapunov
exponents \cite{Kle91}. Thirty years later still, two of us developed an
up-dated version of this hierarchy which, in addition to the spectrum of
Lyapunov exponents, also employs first-return map or Poincar\'e section to
distinguishing between the different types of chaos \cite{Let20b}. The present
classification is a direct continuation of these initial contributions whereby
it is our aim to open up the {\it chaos taxonomy} even wider. Taxonomy is the
science of classification of a given class of objects using a specific
methodology.  The aim of our taxonomy is to provide a classification of chaotic
attractors into different classes allowing to assign each attractor to one of
the pre-established classes. We here means a ``systematic" classification,
that is, the design and utilization of taxonomic ranks to define the different 
classes. This requires a terminology based on essential properties rationally 
deduced \cite{Ere00} that we will here introduce in the context of dynamical
system. This should help to better discriminate whether an attractor is 
actually of a new type or not \cite{Spr11,Let12}. According to this taxonomy, 
an attractor would be new when it is not equivalent to another one at the 
deepest rank. A first impression could be that, asking for a deep taxonomic 
rank could promote claim for new attractor but, most likely, this will be the 
opposite because it would require a deep analysis (too often lacking) showing
that the attractor have an already known topology. In other words, a
non-usual shape of attractor is not sufficient for claiming that this is a new 
one.

We limit ourselves to the case of continuous dynamical systems. This choice is
justified by the fact that a first-return map or a surface of section of a
continuous dynamical system leads to a discrete map. As already mentioned,
the first ingredient for a chaos taxonomy is the number of positive and null
Lyapunov exponents the number of negative Lyapunov exponents
being irrelevant. Nevertheless, despite their usefulness in determining the
dimension, there are also some limitations in only using the spectrum of
Lyapunov exponents, as it will be demonstrated with some examples.

In fact, while developing further this taxonomy, it appeared that the Lyapunov
exponents are the first taxonomic rank, that is, a not so discriminating 
level. We therefore inserted in our taxonomy other ingredients, mainly 
inherited from topology, allowing to refine, step by step, the characterization 
of chaotic attractors. The subsequent part of this paper is organized as 
follows. Section \ref{taxprop} briefly introduces the concepts (Lyapunov 
exponents, bounding tori, first-return maps and branched manifolds) that we 
will use to fully determine a taxon. Section \ref{third} provides some examples of the various taxa which can be observed in three- and four-dimensional state 
spaces, respectively. Section \ref{fourth} gives a full characterization of 
the R\"ossler and Lorenz attractors (with an extension of the description of 
branched manifold by linking matrix). Section \ref{conc} gives some conclusions 
and perspectives to this work.

\section{Essential properties for defining taxa}
\label{taxprop}

The dynamical taxonomy is here proposed for dynamical systems governed by a 
set of $d$ ordinary differential equations
\begin{equation}
  \label{dynsys}
  \dot{\mb{x}} = \mb{f} (\mb{x})
\end{equation}
where $\mb{x} \in \mathbb{R}^d$ is the state vector and $\mb{f} (\mb{x})$ is a
vector field which can be either fully continuous or with a piecewise linear 
switch function, for instance. Since every non-autonomous system can be 
rewritten as an autonomous system \cite{Men00}, there is no need to have 
special consideration for them.

In a talk offered at the University of Utah in 1978, Otto E. R\"ossler
suggested a ``hierarchy theorem'' in the following way \cite{Ros78b}:
\begin{theor}
  \label{theoros}
Every further dimension carries a less trivial type of attractor.
\end{theor}

\begin{quote}
\begin{itemize}

\item $\mathbb{R}^1$: point attractor

\item $\mathbb{R}^2$: periodic attractor

\item $\mathbb{R}^3$: chaotic attractor

\item $\mathbb{R}^4$: hyperchaotic attractor

\item $\mathbb{R}^5$: hyper 2 chaotic attractor

\item $\mathbb{R}^d$ when $d \rightarrow \infty$ limit attractor with
	$(d - 2)$ positive Lyapunov exponents.

\end{itemize}
\end{quote}
He also distinguished {\it minimal} chaos, with a single
positive Lyapunov exponent (for $d > 3$), from {\it maximal} (or full) chaos 
with $d-2$
positive Lyapunov exponents \cite{Ros83}. The number $m$ of positive Lyapunov
exponents is indeed an essential property for (roughly) quantifying the degree 
of chaoticity of a given attractor. Let us designate the $m$ dimensions 
associated with the $m$ positive Lyapunov exponents by C$^m$, ``C'' meaning 
chaotic. Thus C$^0$ is related to periodic or quasi-periodic behavior, C$^1$ to 
chaotic behavior, C$^2$ to hyperchaos, and so on.

One of the missing ingredients in the theorem \ref{theoros} is the role played
by the oscillating nature of the behavior, that is, by tori. In a 
$d$-dimensional state space $\mathbb{R}^d$, it is possible to
embed a torus $T^{n}$ characterized by the product of $n \leq d-1$ circles. A 
torus $T^{d-1}$ is the maximal regular structure which can be embedded in a 
$d$-dimensional space. By regular, we mean a structure which is not
{\it strange} in the sense of 
\cite{Rue71}, that is, which is neither fractal nor leading to a solution which
is sensitive to initial conditions. According to the
distinction between {\it strange} and {\it chaotic} invariant set introduced by 
\cite{Gre84}, we prefer to use the term chaotic than strange for 
non-regular invariant set. In this approach, a point is a torus T$^0 \subset 
\mathbb{R}^1$ and a limit cycle is a torus T$^1 \subset \mathbb{R}^2$. With 
the first dimension occurs the notion of {\it singular point}. With the second 
dimension emerges the notion of {\it oscillation}; the solution can be 
aperiodic (an expanding spiral) in a two-dimensional space, but there is a 
fundamental property which cannot be fulfilled in that case: recurrence, that 
is, boundedness. Without it, the solution does not belong to an invariant set 
and is not further considered here. The present taxonomy is therefore for 
recurrent solutions in the sense of \cite{Bir27}. In a three-dimensional space, 
a torus T$^2$ can be embedded and the corresponding behavior is the so-called 
quasi-periodic one. Such invariant set is
neither strange nor chaotic: it is regular and corresponds to C$^0$. It will be
designated by C$^0$T$^2$. A torus T$^p$ has $p$ null exponents: they correspond 
to the second part of the spectrum of Lyapunov exponents which is here used to 
define a new taxonomic rank. C$^0$T$^1$ and C$^0$T$^2$ are two different taxa, 
distinguished by counting the positive and null Lyapunov exponents: they are
associated with $(m,p) = (0,1)$ and $(0,2)$, respectively. 

Minimal chaos is characterized by one positive Lypaunov exponent: it is 
therefore C$^1$T$^1 \subset \mathbb{R}^d$ with $d > 3$: chaos is necessarily 
maximal in $\mathbb{R}^3$. Recurrent oscillations are 
designated by T$^1$ (one null exponent). A torus T$^n$ is necessarily 
embedded within a space $\mathbb{R}^d$ with $d \geq n +1$. In these two cases, 
the minimal dimension $\delta$ in which those solutions can be observed is such 
as $\delta = m + p + 1$: let us name this minimal dimension, the {\it dynamical 
dimension}. Typically, when there is a single negative Lyapunov exponent, this 
is the dimension $d$ of the system.
There is one special case to discuss for which this relationships does not
work. In its simplest form, toroidal chaos corresponds to a torus whose surface 
is stretched and folded as described by \cite{Cur78}: it is designated by 
C$^1$T$^2$ and one could expect two null and one positive Lyapunov exponents.  
Nevertheless, such a toroidal chaos can be produced by a three-dimensional 
systems for which there is necessarily a negative Lyapunov exponent, imposing
$m+p=2$. In fact, the surface of the torus is not regular since it is stretched 
and folded: the second null exponent which should be associated with the 
toroidal surface is therefore degenerated and is ``merged'' with the positive 
Lyapunov exponent. Three-dimensional toroidal chaos is therefore characterized 
by one positive and one null Lyapunov exponents. Such a special case is 
designated as C$^1$T$^2_{\rm d}$, 
``d'' meaning that there are actually $p-1$ (and not $p$) null exponents: in 
this case, one should speak of ``degenerated toroidal chaos''. We will show 
that when produced by a four-dimensional system, such behavior is no longer
degenerated and two null Lyapunov exponents are actually found as exemplified 
with the driven van der Pol system investigated by \cite{Ued93}) in Section
\ref{third}.

\begin{defi}
A dynamics is said to be {\it degenerated}  when a null Lyapunov exponent is 
merged with a positive one.
\end{defi}

The first two taxonomic ranks are provided by the integers $m$ and $p$,
respectively.

\begin{defi}
The chaotic level is maximal when $m = d-2$ with $d \geq 3$.
\end{defi}

\begin{defi}
The toroidal level is maximal when $p = d-1$ with $d \geq 2$.
\end{defi}

It is thus possible to discriminate some behaviors as follows 
\cite{Kle91,Let20b,Sta20b}.

\begin{center}
  \begin{tabular}{clcl}
    C$^0$T$^1$ & Periodic orbit & C$^2$T$^1$ & Hyperchaos \\
	  C$^0$T$^2$ & Two-frequency quasi-periodic regime & 
	  C$^2$T$^2$ & Toroidal hyperchaos \\
    C$^1$T$^1$ & Chaos & C$^2$T$^2$ & Three-frequency quasiperiodic regime \\
	  C$^1$T$^2$ & Toroidal chaos  & $\hdots$
  \end{tabular}
\end{center}

A dimension connected to the spectrum of Lyapunov exponents is the Kaplan-Yorke 
dimension which is defined as follows \cite{Kap79,Fre83}.

\begin{defi}
Let $\lambda_1 \geq \lambda_2 \geq ... \geq \lambda_j \geq ... \geq \lambda_n$
be the spectrum of ordered Lyapunov exponents and let $k$ be the index such
as $\displaystyle \sum_{i= 1}^k \lambda_i \geq 0$ and
$\displaystyle \sum_{i= 1}^{k+1} \lambda_i < 0$. The Kaplan-Yorke dimension is
\[ D_{\rm KY}
	= k + \frac{\displaystyle \sum_{i=1}^k \lambda_i}
	{\left| \displaystyle \lambda_{k+1} \right|} \, .
\]
\end{defi}

\begin{proper}
By definition, $ k \geq m+ p -1$ for non-degenerated torus, and $k \geq m + p$
otherwise.
\end{proper}

\begin{proper}
By definition, a taxon has a Kaplan-Yorke dimension such as $m + p -1 \leq
D_{\rm KY} \leq d$ for non-degenerated cases, and $m + p - 2 \leq D_{\rm KY}
\leq d$ otherwise.
\end{proper}
It is possible to define the {\it dynamical} dimension $\delta$ as follows.

\begin{defi}
The {\it dynamical dimension} $\delta$ is such as
\begin{equation}
  \delta =
  \left|
    \begin{array}{lcl}
	    m+p & \mbox{ if } & {\rm degenerated}  \\
	    m+p+1 & & {\rm otherwise}
    \end{array}
  \right.
\end{equation}
and $\delta \leq d$. It corresponds to the minimal number of variables required 
to observe a given dynamics.
\end{defi}
The case of {\it conjugated dynamics} will be described in Section 
\ref{third}.

There is a simple relationship between the dynamical dimension $\delta$ and
the Kaplan-Yorke dimension $D_{\rm KY}$: it reads as
\begin{equation}
  \delta = 
  \left|
    \begin{array}{lcl}
       \mbox{Int } (D_{\rm KY}) + 1 & \mbox{ for } & 
	    \mbox{ dissipative system} \\
       D_{\rm KY} & & \mbox{ conservative system.}
    \end{array}
  \right.
\end{equation}
When there is a single negative Lyapunov exponent, $\delta = d$: in other
terms, $\delta = d +1 - n_{\rm n}$ where $n_{\rm n}$ is the number of negative
Lyapunov exponents.

\begin{defi}
The dynamics is maximal when $\delta = d$.
\end{defi}
A maximal dynamics actually explores all the dimensions of the state space: it 
cannot be embedded within a space with a smaller dimension.

\begin{coro}
Dissipative $m$-chaos C$^m$T$^1$ is maximal when $m = d -2$, and is 
necessarily maximal in a three-dimensional space.
\end{coro}

\begin{figure}[ht]
  \centering
  \includegraphics[width=0.90\textwidth]{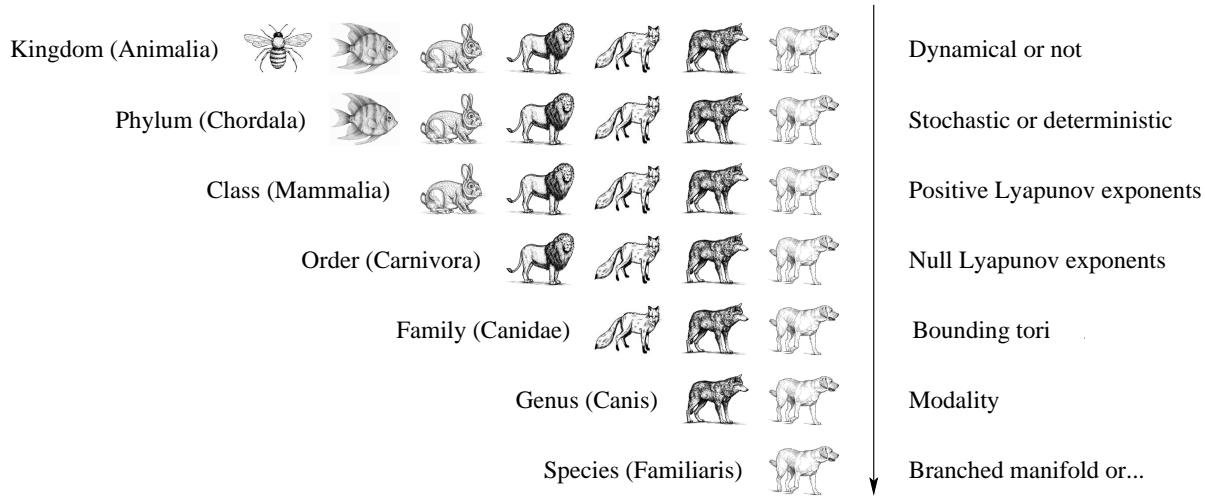} \\[-0.2cm]
  \caption{Analogy between the hierarchy of taxonomic ranks as developed for
animals and those required for fully describing chaotic attractors. The 
``modality'' corresponds to the number of mixing processes. The highest 
taxonomic rank is associated with branched manifold for strongly dissipative 
system whose dynamical dimension is $\delta = 3$.} 
  \label{taxomic}
\end{figure}

Characterizing a chaotic attractor using its spectrum of Lyapunov exponents
corresponds to the first two taxonomic ranks quantified by $m$ and $p$, 
respectively. In the analogy with the taxonomy for classifying animals, these 
two integers could correspond to the {\it class} and the {\it order}
(Fig.\ \ref{taxomic}): for instance, the first two taxonomic ranks for red fox 
is {\it mammal} and {\it carnivore}, 
respectively. The next step would be to frame a little bit more the nature of 
the chaotic invariant set by providing the ``{\it family}'', the analog of 
claiming that red fox is a canidae. 

\begin{figure}[ht]
  \centering
  \begin{tabular}{ccc}
    \includegraphics[width=0.10\textwidth]{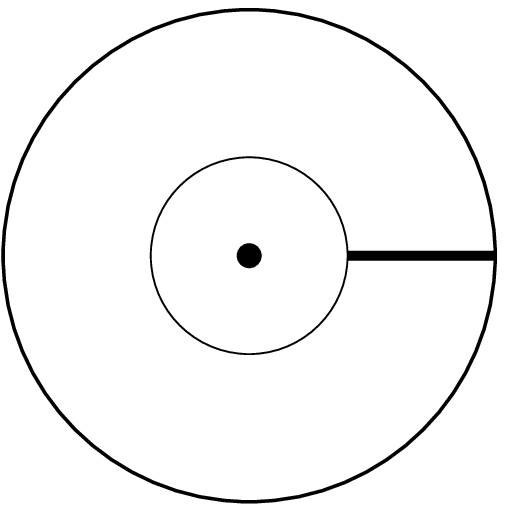} &
    \includegraphics[width=0.22\textwidth]{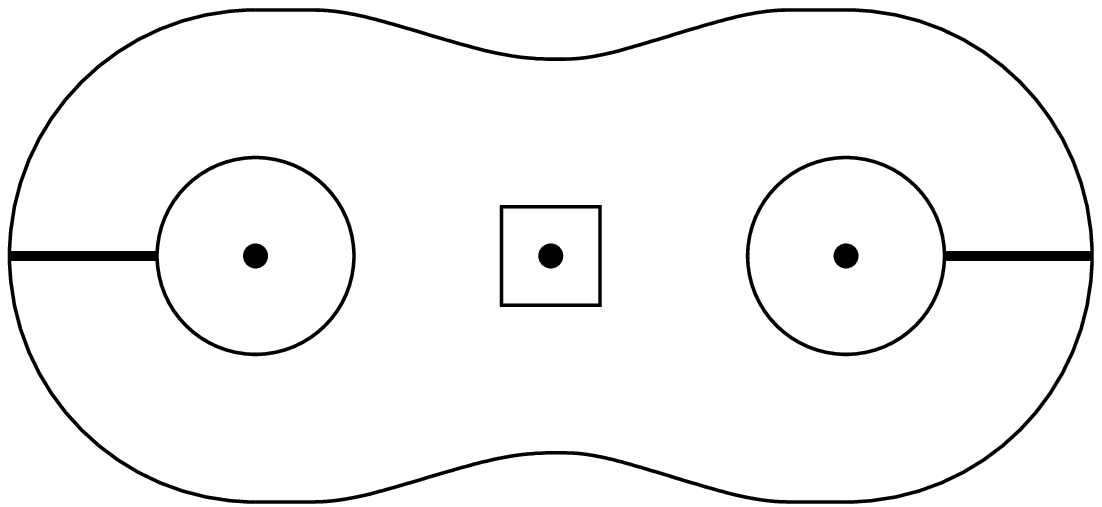} \\
	  {\small (a) $(g,f) = (1,1)$} & {\small (b) $(g,f) = (3,2)$} \\[0.2cm]
    \includegraphics[width=0.22\textwidth]{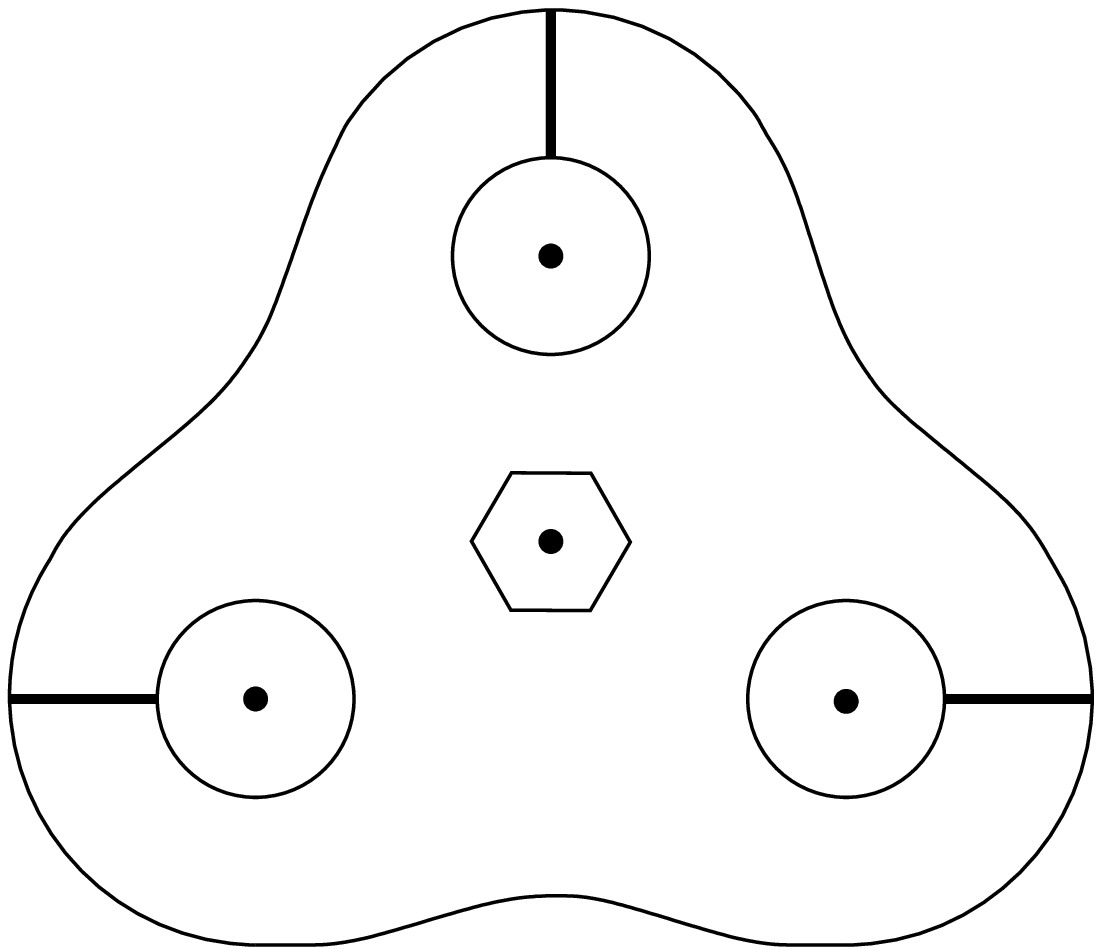} &
    \includegraphics[width=0.22\textwidth]{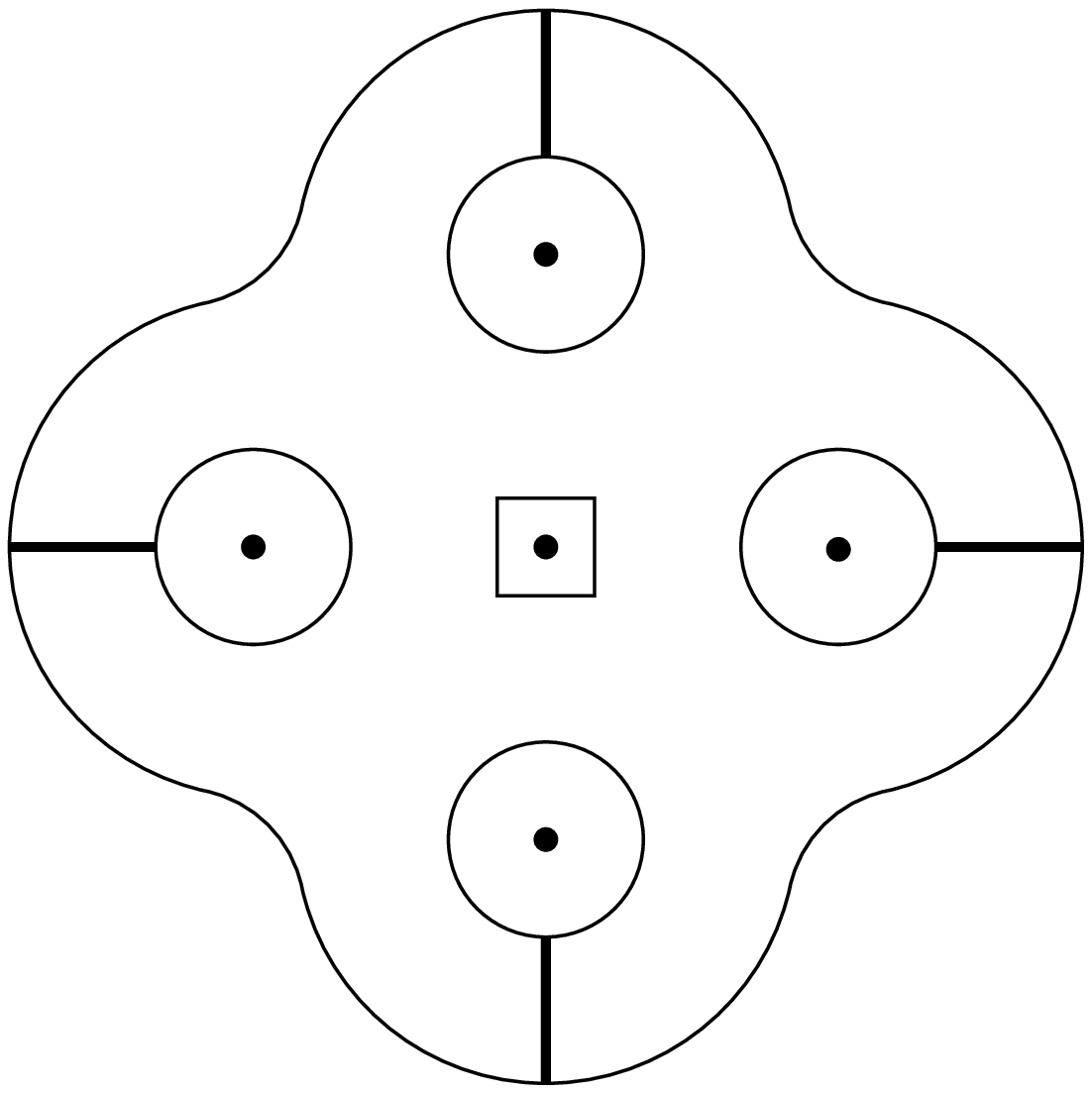} \\
	  {\small (c) $(g,f) = (4,3)$} & {\small (d) $(g,f) = (5,4)$} \\[0.2cm]
    \multicolumn{2}{c}{\includegraphics[width=0.38\textwidth]{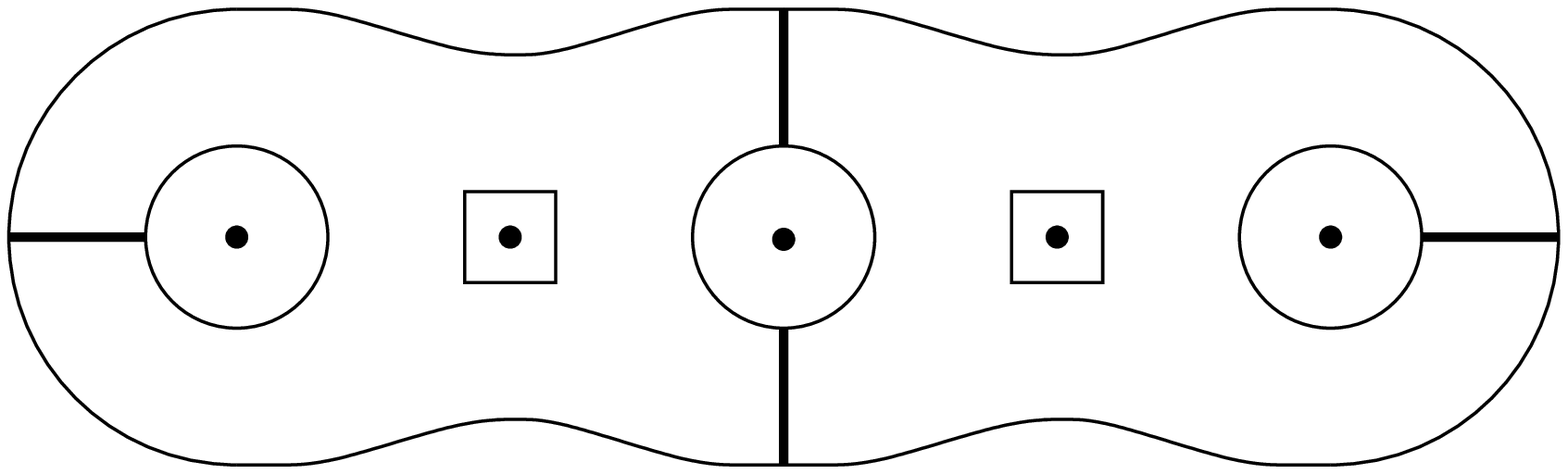}} \\
	  \multicolumn{2}{c}{\small (e) $(g,f) = (5,3)$} \\[-0.2cm]
  \end{tabular}
  \caption{Bounding tori of various genus $g$ ($g \leq 5$). The $g-1$
components of the Poincar\'e section are plotted as thick lines. Case (a) 
applies for the R\"ossler attractor and (b) for the Lorenz attractor. Cases (c) 
and (d) may correspond to the 3- and 4-fold covers of the proto-Lorenz 
system \cite{Mir93,Let95d}. Case (e) can correspond to a multi-scroll 
attractor \cite{Azi99}.}
  \label{boutos}
\end{figure}

For chaotic attractors, there is a concept 
for that: the bounding tori introduced by \cite{Tsa03,Tsa04}. The idea is to 
bound chaotic invariant set by a torus. The corresponding torus has a genus $g$ 
defined by the number of its holes. A few examples of bounding tori are 
provided in Fig.\ \ref{boutos}. It appears that the genus is not sufficient to 
distinguish all possible configurations: for instance, we drew the two 
different bounding tori of genus 5 [Figs.\ \ref{boutos}(d) and 
\ref{boutos}(e)]. At least, a second number should be provided as the number 
$f$ of focus holes around which the trajectory flows. Thus the two genus-5 
bounding tori are characterized by the pairs (5,3) and (5,4), respectively. A 
more sophisticated and complete labelling can be found in \cite{Tsa04}. To 
provide some well-known examples, the spiral R\"ossler attractor is bounded by 
the torus (1,1), the Lorenz attractor is bounded by the torus (3,2), the 
torus (4,3) corresponds to a three-fold cover of the R\"ossler attractor 
\cite{Let01} and a three-fold cover of the proto-Lorenz attractor \cite{Mir93}
is bounded by the torus (4,3).

\begin{figure}[ht]
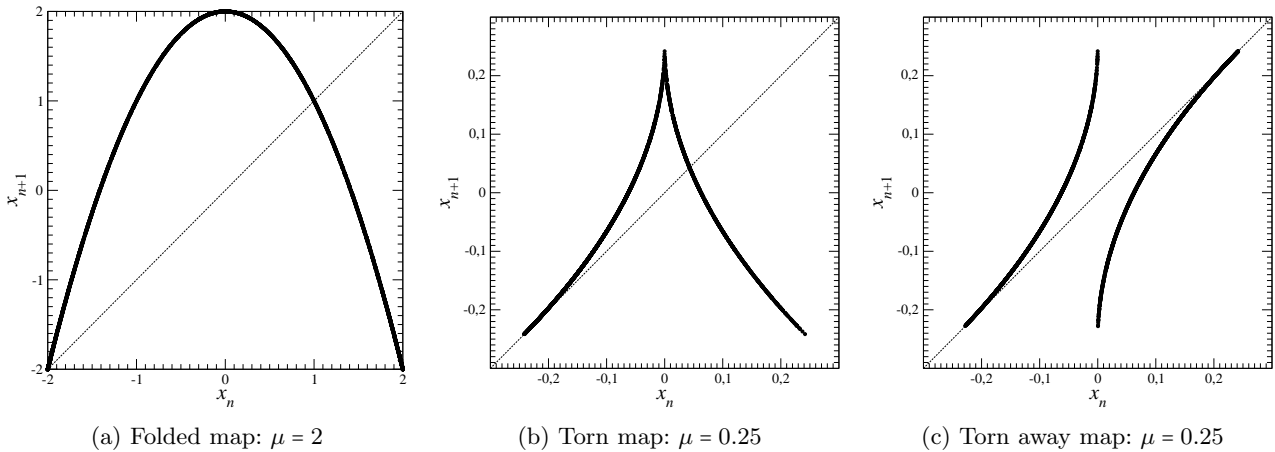

  \centering
  \begin{tabular}{ccc}
	  \includegraphics[width=0.30\textwidth]{folded.eps} & 
	  \includegraphics[width=0.30\textwidth]{torn.eps} &
          \includegraphics[width=0.30\textwidth]{tornaway.eps} \\[0.0cm]
	  {\footnotesize (a) Folded map: $\mu = 2$} &
	  {\footnotesize (b) Torn map: $\mu = 0.25$} &
	  {\footnotesize (c) Torn away map: $\mu = 0.25$} \\[-0.1cm]
  \end{tabular}
  \caption{The three possible unimodal maps.}
  \label{unimaps}
\end{figure}

Once the genus-$g$ bounding torus is determined, the next step is to compute a
first-return map to a ($g-1$)-component Poincar\'e section (when $g >2$). This
means for instance that the Lorenz attractor, which is bounded by a genus-3
torus, must be investigated by using a two-component Poincar\'e section,
leading to a four-branches maps \cite{Let94a,Byr04} as detailed in Section
\ref{fourth}. The description of the different types of first-return map is 
still in its infancy. R\"ossler started to provide various examples of them 
\cite{Ros79b}. The cases of the three possible
unimodal maps (Fig.\ \ref{unimaps}) were extensively treated by \cite{Let21d}. 
The simplest way to describe a first-return map would be to count the number of 
critical points (demarcating the monotone branches) or the number of foldings 
counted in a Poincar\'e section when it is possible to do so. Describing simply 
every type of maps is still an open problem, particularly for weakly 
dissipative systems. Produced by a strongly dissipative system, the spiral 
R\"ossler attractor has one critical point associated with a folding
[Fig.\ \ref{boutos}(b)]. The Lorenz attractor has a first-return map to a 
2-component Poincar\'e section characterized by 3 critical points associated 
with a tearing \cite{Byr04}. Let us here simply use the number $n_{\rm c}$ of 
mixing processes which either corresponds to the number of critical points 
(equal to the number of monotone branches minus one), or the number of foldings 
counted in the Poincar\'e section when toroidal chaotic attractor is observed.
We name {\it modality} the number $n_{\rm c}$.

The last taxonomic rank, corresponding to the most selective one, would be,
for three-dimensional dissipative systems, a branched manifold 
\cite{Min92,Tuf92,Let95a,Gil98}. The simplest branched manifolds associated with
the three types of map shown in Fig.\ \ref{unimaps} are discussed 
by \cite{Let21d}. We postpone for future works to provide this 
last taxonomic rank for all the systems investigated in the next section,
the problem being still open for some cases.

Using all these essential properties considered as different taxonomic ranks,
a taxon could be described by a set of integers as follows.

\begin{quote}
\begin{itemize}

\item the number $m$ of positive Lyapunov exponents;

\item the number $p$ of null Lyapunov exponents;

\item the genus $g$ of the bounding torus and its number $f$ of focus holes
(Betti numbers could be an efficient way to describe them \cite{Cha20,Cha21});

\item the modality $n_{\rm c}$, that is, the number of critical points or of 
mixing processes (folding or tearing);

\item the last taxonomic rank is not yet sufficiently advanced to be fully
summarized by a set of integers. Attractors with a dynamical dimension 
$\delta = 3$ can be described by a branched manifold \cite{Gil98}: when they
are bounded by genus-1 torus, they can be described by a 
$n_{\rm c} \times n_{\rm c}$ linking matrix \cite{Tuf92,Let95a}. The case of 
higher genus $g$ tori will be treated in Section
\ref{fourth}. More generalized topological description are still to be 
developed for high-dimensional and/or conservative systems.

\end{itemize}
\end{quote}

Providing these numbers allows to refine each taxon as illustrated in Fig.\
\ref{taxomic}. It is worthy to note that in most of the literature, the
attractors are only described using the first taxonomic ranks, that is, the 
less discriminating ones. The taxon associated with a given chaotic attractor 
is fully described only when the last taxonomic rank is determined,
that is, when a branched manifold (or some generalized topological 
description) is provided.

\section{Examples of various types of invariant sets}
\label{third}

To help the reader to have some intuition with our dynamical taxonomy, we now 
propose a set of different types of behaviours which can be 
produced by three- and four-dimensional systems. The invariant sets are plotted
with some of their characteristics in Table \ref{dytaxa}: they are produced
by the systems which are reported with the retained parameter values in Table
\ref{dyneqs}. All our examples are such that 
there is a single negative Lyapunov exponent, in agreement with our assumption 
that additional negative Lyapunov exponents are irrelevant for the taxonomy of
invariant sets. Consequently, the dynamical dimension $\delta$ is here equal to
the dimension $d$ of the system. Cases (a), (b), (e), (g), (h), and (j)
are non-degenerated cases where $m$ ($p$) is the number of positive (null) 
Lyapunov exponents. All the invariant sets but cases (c) and (h) are bounded by 
a trivial genus-1 torus: a single component is therefore required for computing 
the Poincar\'e section. Case (a) is a quasi-periodic regime C$^0$T$^2$ as 
revealed by the annular Poincar\'e section [Table \ref{dytaxa}(a)] and there is 
no mixing process ($n_{\rm c} = 0$): the modality is null. Case (b) is a 
maximal chaos characterized by a unimodal map ($n_{\rm c} = 1$) produced by the
R\"ossler system. Case (e) is a 
toroidal chaos C$^1$T$^2 \subset \mathbb{R}^4$ produced by the mixed driven 
van der Pol system with five foldings [they can be counted from the inner 
part of the Poincar\'e section as indicated by the red disks in Table\ 
\ref{dytaxa}(e)]: the modality is thus $n_{\rm c} = 5$. This latter system is 
mixed because it is made of dissipative van der Pol oscillator driven by a 
conservative harmonic oscillator \cite{Men00}.

Case (g) is a toroidal chaos C$^1$T$^2 \subset \mathbb{R}^4$ produced by the
conservative H\'enon-Heiles system. In such a case, $D_{\rm KY} = 4$. The
chaotic sea is bounded by a genus-1 torus [Table\ \ref{dytaxa}(g)]; it is 
possible to refine the bounding of the invariant set by associating tori with 
the different quasi-periodic islands exhibited in the Poincar\'e section. There 
is at least five additional tori, more or less knotted inside the main genus-1 
bounding torus, which can be used to bound the chaotic sea. They are relevant 
to characterize the structure of conservative toroidal chaos \cite{Tri11}.
We therefore suggest to set the modality to $n_{\rm c} = 5$ for this type of 
toroidal chaos. According to the procedure developed by \cite{Man21a}, this 
attractor should be characterized by a seven-strip branched manifold but this 
is out of the scope of the present paper. 

Case (j) is the first-example of hyperchaotic chaos 
C$^2$T$^1 \subset \mathbb{R}^4$ discussed by \cite{Ros79a}. The first-return 
map to the Poincar\'e section has at least four branches [Table\ 
\ref{dytaxa}(j)]. The presence of a two-dimensional unstable manifold induces
a very thick first-return map which makes the partition difficult to establish.
Within the scope of this work, we will limit ourselves to state that the 
modality is $n_{\rm c} \geq 4$.

Among the five cases remaining to investigate, four are degenerated. Case (c)
is a toroidal chaos whose surface is stretched and folded as depicted in the
Curry-Yorke scenario \cite{Cur78}. Since the surface is stretched, it can no
longer be associated with a null Lyapunov exponent but with a positive one. 
Since the structure is still characterized by an annular structure [Table\
\ref{dytaxa}(c)], it should be designated as C$^1$T$^2$. We thus consider that
the second null Lyapunov exponent is degenerated and is merged with the 
positive one. We have $p-1$ null Lyapunov exponents. There is four foldings
[see the red disks in Table\ \ref{dytaxa}(c)], leading to a modality $n_{\rm c}
= 4$. A six-strip branched manifold should describe the topology of this 
attractor (postponed for future works).

Case (h) is a nearly conservative toroidal chaos --- the averaged trace 
$\overline{{\rm Tr }J} = - 6 \cdot 10^{-5}$ is slightly negative --- produced by
a mixed four-dimensional system. This is an example of invariant set bounded by
a genus-3 torus [see Fig.\ \ref{boutos}(b)]: a property which results from its 
inversion symmetry. There is two sets of quasi-periodic islands and, 
consequently, two inner tori bound the chaotic sea [Table\ \ref{dytaxa}(h)]. 
Compared to the chaotic sea produced by the H\'enon-Heiles system [Table\ 
\ref{dytaxa}(g)], it is simpler since it presents a smaller number of 
quasi-periodic islands: using the conjecture used for the H\'enon-Heiles 
system, the modality would be $n_{\rm c} = 2$.

\begin{table}[htbp]
  \tbl{Invariant sets produed by the dynamical systems here investigated. Are 
reported for each one: the type C$^m$T$^p$, the spectrum of Lyapunov exponents, 
the Kaplan-Yorke dimension $d_{\rm KY}$, the dynamical dimension $\delta$, and 
the number $n_{\rm c}$ of mixing processes (critical points, foldings). A plane
projection and a Poincar\'e section or a first-return map [cases (b) and (i)] 
are provided.
	\label{dytaxa}}
 {\begin{tabular}{cccccccc}
   \toprule
	 & State portrait & Map & & State portrait & Map \\[0.1cm] \hline
	 \\[-0.2cm]
	 & 
    \includegraphics[height = 0.16\textwidth]{stank3Da13.eps} &
    \includegraphics[height = 0.16\textwidth]{stankmap3D150.eps} & &
    \includegraphics[height = 0.16\textwidth]{henheiles.eps} &
    \includegraphics[height = 0.16\textwidth]{heilesectiona.eps} 
	 \\[-2.9cm]
     $ \begin{array}{l}
	     \mbox{(a) [1] C}^0{\rm T}^2 \\[0.1cm] 
	     \lambda_1 = 0.0 \\
	     \lambda_2 = 0.0 \\
	     \lambda_3 = -0.0803 \\
	     d_{\rm KY} = 2.0 \\
	     \delta = 3 \\
	     n_{\rm c} = 0 
     \end{array} $ & & & 
     $ \begin{array}{l}
	     \mbox{(g) [6] C}^0{\rm T}^2 \\[0.1cm] 
	     \lambda_1 = 0.0507 \\
	     \lambda_2 = 0.00 \\
	     \lambda_3 = 0.00 \\
	     \lambda_4 = -0.0507 \\
	     d_{\rm KY} = 4.0 \\
	     \delta = 4   \\
	     n_{\rm c} \stackrel{?}{=} 4 \\ 
     \end{array} $ 
	 \\[1.5cm] \hline \\[-0.2cm]
	 & 
    \includegraphics[height=0.16\textwidth]{rosatco.eps} &
    \includegraphics[height = 0.16\textwidth]{rosmapco.eps} & & 
    \includegraphics[height = 0.16\textwidth]{dougyre.eps} & 
    \includegraphics[height = 0.16\textwidth]{dougyrsecder.eps} 
	 \\[-2.9cm]
    $\begin{array}{l}
	    \mbox{(b) [2] C}^1{\rm T}^1 \\[0.1cm]
	    \lambda_1 = 0.105 \\
	    \lambda_2 = 0.0 \\
	    \lambda_3 = -3.2779 \\
	    d_{\rm KY} = 2.032 \\
	    \delta = 3 \\
            n_{\rm c} = 1
     \end{array}$ & & & 
     $ \begin{array}{l}
	     \mbox{(h) [7] C}^1{\rm T}^2 \\[0.1cm] 
             \lambda_1 = 0.06579 \\
	     \lambda_2 = 0.0 \\
	     \lambda_3 = 0.0 \\
	     \lambda_4 = -0.06758 \\
	     d_{\rm KY} = 3.97 \\
	     \delta = 4 \\
	     n_{\rm c} \stackrel{?}{=} 2 
     \end{array} $ 
	 \\[1.6cm] \hline \\[-0.2cm]
	 & 
    \includegraphics[height = 0.16\textwidth]{kuzoscko.eps} &
    \includegraphics[height = 0.16\textwidth]{kuzQPmap.eps} & &
    \includegraphics[height = 0.16\textwidth]{kleintorusT3.eps} & 
    \includegraphics[height = 0.16\textwidth]{torusT3sec.eps} 
	 \\[-2.9cm]
    $\begin{array}{l}
		     \mbox{(c) [1] C}^1{\rm T}^2_{\rm d} \\[0.1cm] 
		     \lambda_1 = 0.1408 \\
		     \lambda_2 = 0.0 \\
		     \lambda_3 = -5.1834 \\
		     d_{\rm KY} = 2.0272 \\
		     \delta = 3 \\
		     n_{\rm c} = 4
    \end{array}$ & & & 
     $ \begin{array}{l}
	     \mbox{(i) [5] C}^1{\rm T}^3 \\[0.1cm] 
             \lambda_1 = 0.003 \\
	     \lambda_2 = 0.0 \\
	     \lambda_3 = 0.0 \\
	     \lambda_4 = -0.003 \\
	     d_{\rm KY} = 4.0 \\
	     \delta = 4 \\
	     n_{\rm c} = ?
     \end{array} $ 
	 \\[1.6cm] \hline \\[-0.2cm]
	 & 
    \includegraphics[height = 0.16\textwidth]{jerktorsea.eps} &
    \includegraphics[height = 0.16\textwidth]{jerkcomap.eps} & & 
    \includegraphics[height = 0.16\textwidth]{koC2.eps} &
    \includegraphics[height = 0.16\textwidth]{koC2map.eps} \\
    \\[-3.3cm]
             $\begin{array}{l}
		     \mbox{(d) [3] C}^1{\rm T}^2_{\rm d} \\[0.1cm]
		     \lambda_1 = 0.0179 \\
		     \lambda_2 = 0.0 \\
		     \lambda_3 = -0.0179 \\
		     d_{\rm KY} = 3.0 \\
		     \delta = 3 \\
		     n_{\rm c} = ? 
    \end{array}$ & & & 
     $ \begin{array}{l}
	     \mbox{(j) [8] C}^2{\rm T}^1 \\[0.1cm] 
             \lambda_1 = 0.115 \\
	     \lambda_2 = 0.021 \\
	     \lambda_3 = 0.0  \\
	     \lambda_4 = -24.931 \\
	     d_{\rm KY} = 3.01 \\
             \delta = 4 \\
	     n_{\rm c} \geq 4
     \end{array} $ 
	 \\[1.6cm] \hline \\[-0.2cm]
	 & 
    \includegraphics[height = 0.16\textwidth]{ueda.eps} &
    \includegraphics[height = 0.16\textwidth]{uedamap.eps} & & 
    \includegraphics[height = 0.16\textwidth]{anhypertoro.eps} &
    \includegraphics[height = 0.16\textwidth]{animap809.eps} 
    \\[-2.9cm]
             $\begin{array}{l}
		     \mbox{(e) [4] C}^1{\rm T}^2 \\[0.1cm] 
		     \lambda_1 = 0.5762 \\
		     \lambda_2 = 0.0 \\
		     \lambda_3 = 0.0 \\
		     \lambda_4 = -0.7559 \\
		     d_{\rm KY} = 3.7623 \\
		     \delta = 4 \\
		     n_{\rm c} = 5 
	     \end{array}$ & & & 
     $ \begin{array}{l}
	     \mbox{(k) [9] C}^2{\rm T}^2_{\rm cd} \\[0.1cm] 
	     \lambda_1 = 0.0108 \\
	     \lambda_2 = 0.0005 \\
	     \lambda_3 = 0.0 \\
	     \lambda_4 = -0.138 \\
	     d_{\rm KY} = 3.08 \\
	     \delta = 4 \\
	     n_{\rm c} = 2
     \end{array} $ 
	 \\[1.3cm] \hline \\[-0.2cm]
	 & 
    \includegraphics[height = 0.16\textwidth]{torklein.eps} &
    \includegraphics[height = 0.16\textwidth]{kotorC-T2map.eps} \\[-2.9cm]
             $\begin{array}{l}
		     \mbox{(f) [5] C}^1{\rm T}^2_{\rm c} \\[0.1cm]
		     \lambda_1 = 0.0088 \\
		     \lambda_2 = 0.0 \\
		     \lambda_3 = 0.0  \\
                     \lambda_4 = -0.0588 \\
		     d_{\rm KY} = 3.1492 \\
		     \delta = 4 \\
		     n_{\rm c} = 1 
	     \end{array}$ 
	 \\[1.1cm] 
	 \botrule
	 \\[-0.3cm]
  \multicolumn{6}{p{0.99\textwidth}}{[1] = \cite{Sta20b}, [2] = \cite{Ros76c},
 [3] = \cite{Spr97b}, [4] = \cite{Ued93}, [5] = \cite{Kle91}, [6] = 
 \cite{Hen64}, [7] = \cite{Cha19}, [8] = \cite{Ros79a}, [9] = \cite{Ani05} }
  \end{tabular} }
\end{table}

\clearpage
\begin{table}[htpb]
  \tbl{Governing equations of the dynamical systems here investigated. 
Parameter values and initial conditions (IC) when they may be sensitive) are 
reported.
	\label{dyneqs}}
 {\begin{tabular}{clll}
	 \toprule
	 & Reference & Governing equations & \\[0.1cm]
	 \hline
	 \\[-0.3cm]
	 (a) & \cite{Sta20b} & 
   $\begin{array}{l}
      \dot{x} = y \\[0.1cm]
      \dot{y} = \left( \displaystyle \alpha + z + x^2 - \beta x^4 \right) 
	   \, y - \omega_0^2 \, x \\[0.1cm]
	    \dot{z} = \mu - z - \epsilon y^2
    \end{array} $  & 
   $\begin{array}{ll}
           \alpha = 1.5 & \beta = 0.04 \\
	   \mu = 4 & \epsilon = 0.02 \\
	   \omega_0 = 2 \pi
    \end{array}$ 
	 \\[0.3cm] \hline \\[-0.3cm]
	 (b) & \cite{Ros76c} &
    $\begin{array}{l}
       \dot{x} = -y -z \\[0.1cm]
       \dot{y} = x + ay \\[0.1cm]
       \dot{z} = b + z (x-c)  \, .
     \end{array}$ & 
   $\begin{array}{ll}
	a=0.43295 \\
	b = 2 \\
	c=4
    \end{array}$ 
	 \\[0.3cm] \hline \\[-0.3cm]
	 (c) & \cite{Sta20b} & 
   $\begin{array}{l}
      \dot{x} = y \\[0.1cm]
      \dot{y} = \left( \displaystyle \alpha + z + x^2 - \beta x^4 \right) 
	   \, y - \omega_0^2 \, x \\[0.1cm]
	    \dot{z} = \mu - z - \epsilon y^2
    \end{array} $  & 
   $\begin{array}{ll}
	  \alpha = 9 & \beta = 0.016 \\
	   \mu = 4 & \epsilon = 0.02 \\
	   \omega_0 = 2 \pi
    \end{array}$ 
	 \\[0.1cm] \hline \\[-0.3cm]
	 (d) & \cite{Spr97b} & 
    $\begin{array}{l}
	    \dot{x} = y \\[0.1cm]
	    \dot{y} = z \\[0.1cm]
	    \dot{z} = a (1 - x^2) x - x^2y
    \end{array}$ & 
   $\begin{array}{lcl}
	   a = 0.027 ~~~ & ~ & x_0 = 0.98 \\
	   & & y_0 = 0.58 \, x_0 \\
	   & & z_0 = 0
    \end{array}$ 
	 \\[0.1cm] \hline \\[-0.3cm]
	 (e) & \cite{Ued93} & 
   $\begin{array}{l}
	   \dot{x} = y \\[0.1cm]
     \dot{y} = \mu (1- \gamma x^2) y - x^3 + u \\[0.1cm]
     \dot{u} = v \\[0.1cm]
     \dot{v} = - \omega^2 u
   \end{array}$ &
   $\begin{array}{lcl}
	   \mu=0.2  ~~~~~ & ~ & x_0 =0.2 \\
	   \gamma=8 & & y_0 =0.2 \\
	   B=0.35 & & u_0 = B \\
	   \omega = 1.02 & & v_0 = 0
    \end{array}$ 
	 \\[0.1cm] \hline \\[-0.3cm]
	 (f) & \cite{Kle91} & 
     $\begin{array}{l}
       \dot{x} = y \\[0.1cm]
       \dot{y} = -x-a z-b w \\[0.1cm]
       \dot{z} = d \left( \displaystyle 1-x^2 \right)-c w \\[0.1cm]
       \dot{w} = c z-e w \, .
     \end{array}$ &
   $\begin{array}{lcl}
	   a = 0.15 & ~ & x_0 = -0.043 \\
	   b = 0.25 & & y_0 = -0.157 \\
	   c = 0.1 & & z_0 = -3.113 \\
	   d = 0.3922 & & w_0 = -1.826 \\
	   e = 0.05 
    \end{array}$ 
	 \\[0.1cm] \hline \\[-0.3cm]
	 (g) & \cite{Hen64} & 
    $\begin{array}{l}
      \dot{x} = V_x \\[0.1cm]
      \dot{V}_x = - x - 2 xy  \\[0.1cm]
      \dot{y} = V_y \\[0.1cm]
      \dot{V}_y = - y - x^2 + y^2
    \end{array}$ & 
   $\begin{array}{lcl}
	   ~~~~~~~~~~~~~~~ & ~ & x_0 = 0 \\
	   & & y_0 = -0.15 \\
	   & & V_{x,0} = 0.478 \\
	   & & V_{y,0} = 0
    \end{array}$ 
	 \\[0.1cm] \hline \\[-0.3cm]
	 (h) & \cite{Cha19} & 
    $\begin{array}{rl}
      \dot{x}
      = & A \pi \sin
        \left( \pi \left[ \displaystyle x^2 u + x - u \right]
                      \right) \sin (\pi y) \\[0.1cm]
      \dot{y}
      = & A \pi \cos \left( \pi \left[ x^2 u + x - u \right]
	    \right) \cos (\pi y) \\[0.1cm]
      \dot{u} = & v \\
      \dot{v} = & - \omega^2 u  \, .
    \end{array}$ &
   $\begin{array}{lcl}
	   A = 0.1 & ~~~~~ & x_0 = 0.2 \\
	   \eta = 0.1 & & y_0 = 0 \\
	   \omega = \frac{\pi}{5} & & u_0 = 0 \\
	   & & v_0 = \omega \, \eta 
    \end{array}$ 
	 \\[0.1cm] \hline \\[-0.3cm]
	 (i) & \cite{Kle91} & 
    $\begin{array}{l}
      \dot{x} = y \\
      \dot{y} = -x -z -w \\
      \dot{z} = cw \\
      \dot{w} = a \left( \displaystyle 1-x^2 \right) - cz \, .
    \end{array}$ &
   $\begin{array}{lcl}
	   a = 0.03 & ~ & x_0 = 0.6173 \\
	   c = 0.1313 & & y_0 = -0.1368 \\
	   & & z_0 = -0.0078 \\
	   & & w_0 = -0.1231 
    \end{array}$ 
	 \\[0.1cm] \hline \\[-0.3cm]
	 (j) & \cite{Ros79a} & 
     $\begin{array}{l}
             \dot{x} = y +ax +w \\
             \dot{y} = -x-z \\
             \dot{z} = b+z y \\
             \dot{w} = -c z+d w
     \end{array}$ & 
   $\begin{array}{lcl}
	   a=0.25 & ~~~ & x_0 = -10 \\
	   b=3 & & y_0 = -6 \\
	   c=0.5 & & z_0 =0 \\
	   d=0.05 & & w_0 = 10.1 
    \end{array}$ 
	 \\[0.1cm] \hline \\[-0.3cm]
	 (k) & \cite{Ani05} & 
    $\begin{array}{l}
	    \dot{x} = - y \\
	    \dot{y} = x + \mu y - yw - \delta y^3 \\
	    \dot{z} = w \\
	    \dot{w} = - \beta z - \alpha w + \alpha \Phi (y)
    \end{array}$ &
   $\begin{array}{lcl}
	   \alpha = 0.2 & ~~~ & \\
	   \beta = 0.43 & & \\
	   \delta = 0.001 & & \\
	   \mu = 0.0809 & & \\
	   \theta = \frac{\pi}{3} & & 
    \end{array}$ 
	 \\[0.1cm] \botrule
  \end{tabular} }
\end{table}

\begin{figure}[ht]
  \centering
    \includegraphics[width=0.33\textwidth]{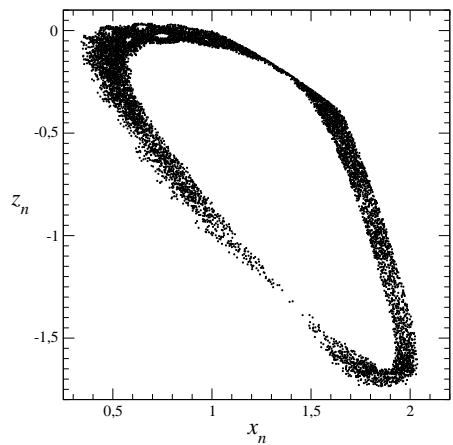} \\
  \caption{Second-order Poincar\'e sections ${\cal P}_1$ and
${\cal P}_2$ of the toroidal chaos C$^1$T$^3$ produced by the four-dimensional
conservative oscillator (i). Same parameter values as reported in Table\
	\ref{dyneqs}(i) and $\epsilon = 0.04$.}
  \label{torT3sea2}
\end{figure}

Case (i) is another example of toroidal chaos produced by a conservative
system: rather than governed by two frequencies as in the cases (c) and (e),
there are three different frequencies involved in the toroidal structure. This 
is therefore a toroidal chaos C$^1$T$^3_{\rm d} \subset \mathbb{R}^4$. It is 
degenerated for reasons similar to those involved in the case (c). The 
Poincar\'e section [Table\ \ref{dytaxa}(h)] looks like a torus: this is a
first-order Poincar\'e section (as all the other sections discussed in this 
work) defined as
\begin{equation}
  {\cal P}_1 \equiv \left\{ \displaystyle (x_n,z_n, w_n) \in \mathbb{R}^3 ~|~
	y_n = 0 \right\}  \, .
\end{equation} 
To exhibit the toroidal nature of this invariant set, a second-order Poincar\'e 
section can be used \cite{Par89}: it is defined as
\begin{equation}
  {\cal P}_2 \equiv
  \left\{ \displaystyle
    \left( x_n, z_n \right) \in \mathbb{R}^2 ~|~
        y_n = 0, | w_n | < \epsilon, \dot{w}_n > 0
  \right\}
\end{equation}
and corresponds to the intersections of the trajectory with the first-order
Poincar\'e section and located in a small domain here defined by the 
red dashed lines in Table\ \ref{dytaxa}(i). The thickness $\epsilon$ is set as 
a compromise between the accuracy of the plot and the time required to compute 
it: with $\epsilon = 0.04$, to get 32,0000 points in ${\cal P}_2$, we integrated
the system over $8.4 \cdot 10^8$ time steps, leading to $6.4 \cdot 10^5$ 
intersections with ${\cal P}_1$. An annular shape is clearly revealed by the
section ${\cal P}_2$ (Fig.\ \ref{torT3sea2}). This behavior is degenerated 
since the torus T$^3$ is characterized by two null Lyapunov exponents and 
not three.

The last degenerated case, system (k), is discussed with the case (f). Let us
start with the latter. The Poincar\'e section has the typical annular structure of a toroidal behavior [Table\ \ref{dytaxa}(f)]. Nevertheless, the section 
looks like a spiral R\"ossler attractor. In fact, this behavior is observed 
after a period-doubling cascade in tori when parameter $d$ is varied. Such a 
route to chaos was reported by \cite{Fra83,Kan83,Arn83,Kle91,Let07a}. It was 
shown that such a behaviour can be produced by two conjugated 
maps \cite{Kan86}. The specificity of this toroidal chaos arising after a 
period-doubling cascade on tori is that it cannot be degenerated in a 
three-dimensional space. Indeed, to have a ``horseshoe map'' conjugated to a 
toroidal structure, it is necessary to have a four-dimensional space, exactly
as it is required to have such a space for embedding a Klein bottle 
\cite{Kle82}. A Klein bottle is a closed, single-sided mathematical surface of 
genus 2 \cite{Seq13}: it is the fourth-dimensional version of the 
three-dimensional M\"obius band. It is not possible to 
construct a Klein surface without self-intersection in a three-dimensional 
space: the fourth dimension is required for that. For similar reason, it is not 
possible to get conjugated toroidal chaos CT$^2_{\rm c}$ in a three-dimensional 
space without self-intersection. Conjugated toroidal chaos C$^1$T$^2_{\rm c} 
\subset \mathbb{R}^4$ is thus structurally different from the toroidal chaos 
C$^1$T$^2_{\rm d}$ or the toroidal chaos C$^1$T$^2$ --- produced by the 
oscillator investigated by \cite{Sta20b} and by the driven van 
der Pol system \cite{Ued93}, respectively --- which results from a Curry-Yorke 
scenario and can be embedded within a three-dimensional space. 

\begin{figure}[ht]
  \centering
  \begin{tabular}{cc}
    \includegraphics[width=0.17\textwidth]{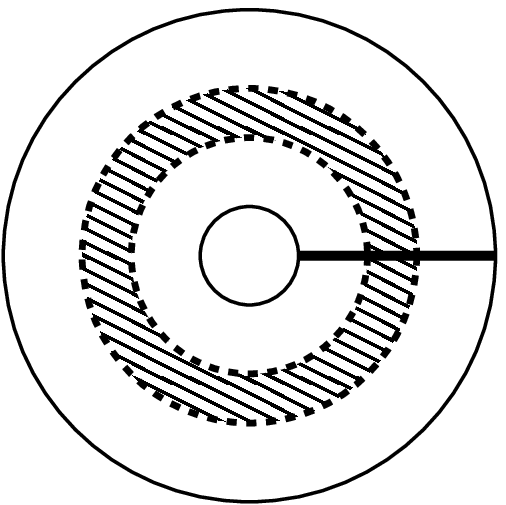} &
    \includegraphics[width=0.17\textwidth]{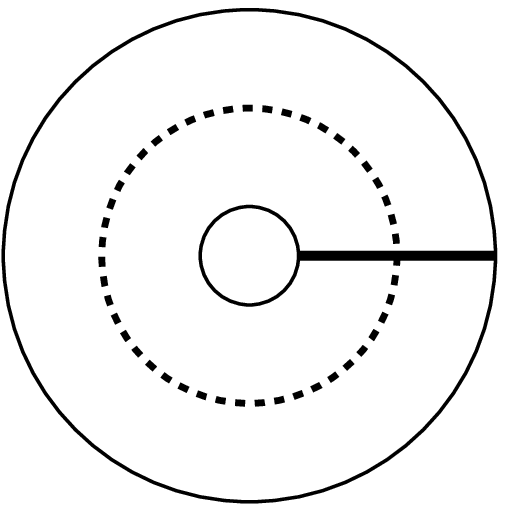} \\
	    {\footnotesize (a) Just after the period-doubling} & 
	    {\footnotesize (b) Homoclinic situation} \\[-0.2cm]
  \end{tabular}
  \caption{Bounding tori drawn in a Poincar\'e section for the conjugated 
toroidal chaos C$^1$T$^2_{\rm c}$ produced by the system (k). (a) Just after 
the period-doubling cascade in tori, the invariant set is bounded by one outer 
torus (solid lines) and one inner torus (dashed lines) which defines the 
forbidden domain (hatched domain). (b) In the homoclinic situation, the inner 
bounding torus is reduced to a singular circle (dashed circle).}
  \label{genuntor}
\end{figure}

Case (k) is quite similar to case (f) but it presents a possibility which is 
not observed in the system (f): as observed in the R\"ossler system, after the
period-doubling cascade, the attractor is developed up to a homoclinic 
situation \cite{Let95a,Maly20} where the attractor is bounded with a torus 
whose inner radius is null [Fig.\ \ref{genuntor}(b)]. The Poincar\'e section
does no longer present a non-visited domain in its centre. Then the fourth
dimension allows a toroidal hyperchaos C$^2$T$^2_{\rm d}$ [Table\
\ref{dytaxa}(k)].

\begin{figure}[ht]
  \centering
	\includegraphics[width = 0.35\textwidth]{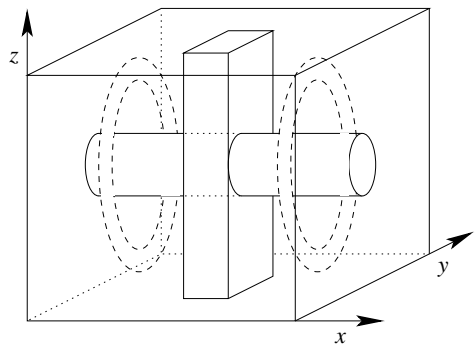} \\[-0.2cm]
  \caption{Bounding torus for the chaotic sea produced by the conservative jerk
system (g). The two holes are organized as a crux: this is a genus-3
bounding torus. There is two inner trivial bounding tori (dashed lines).
Parameter value: $a = 0.027$.}
  \label{jerkbotor}
\end{figure}

The last case is produced by the jerk conservative system introduced by 
\cite{Spr97b}. It is a metastable chaos since the invariant set is bounded by 
an unstable manifold connected to the origin of the state space which ejects
the trajectory to infinity. This is a toroidal chaos C$^1$T$^2_{\rm d}$, of a
type which is similar to the toroidal chaos C$^1$T$^2$ produced by the 
H\'enon-Heiles system [Table\ \ref{dytaxa}(g)], but here degenerated. Another
particularity, possibly imposed by the inversion symmetry, is that the 
invariant set is bounded by a non-trivial genus-3 torus, the focus hole 
($y=z=0$) crossing the saddle hole ($x=y=0$) (Fig.\ \ref{jerkbotor}). Showing
that such a bounding torus is of genus 3 is detailed by \cite{Let09c}. This 
type of chaos must be therefore be investigated with a 2-component Poincar\'e
section as plotted in Table\ \ref{dytaxa}(d).

\section{Complete treatment of the R\"ossler and Lorenz attractors}
\label{fourth}

\begin{figure}[ht]
  \centering
	\includegraphics[width=0.30\textwidth]{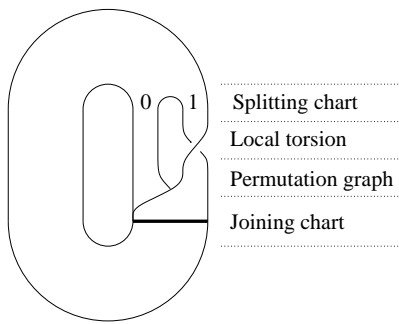} \\[-0.2cm]
  \caption{The branched manifold corresponding to the spiral R\"ossler 
	attractor plotted in the space $\mathbb{R}^3 (X,Y,Z)$.}
  \label{bramafold}
\end{figure}

In this Section, we fully describe three different taxa. Let us start with the 
R\"ossler spiral attractor as plotted in Table\ \ref{dytaxa}(b). The spiral 
R\"ossler attractor is topologically described by the branched manifold shown 
in Fig.\ \ref{taxomic}(a) when the R\"ossler system is rewritten in the space 
$\mathbb{R}^3 (X,Y,Z)$ with $X = y$, $Y = x$, and $Z = z$ \cite{Let21d}. This 
branched manifold is drawn according to the standard insertion introduced by 
\cite{Tuf92}: they are joined from the back to the front, and from the left to 
the right (Fig.\ \ref{bramafold}). This avoids to use a second matrix to encode 
the insertion as used by \cite{Gil98,Gil03}. Strips are labelled 0 and 1, from 
the center of the attractor to its periphery and according to the natural order 
\cite{Bai89}. The symbols are also chosen with respect to the parity of the 
strip: an even (odd) integer for an order preserving (reversing) strip. The 
strips are joined and squeezed at the thick line corresponding to the ideal 
location of the Poincar\'e section of the attractor. 

The part of the branched manifold between the splitting chart and the joining 
chart is called a {\it linker} \cite{Ros13} and can be described by a linking 
matrix reading as \cite{Tuf92,Let95a,Ros13}
\begin{equation}
  \label{linker}
  {\cal L}_{\rm R} = 
  \left[
    \begin{matrix}
      0 & 0 \\
      0 & + 1
    \end{matrix}
  \right\rsem
\end{equation}
where element $L_{ii}$ corresponds to the local torsion (the number of 
$\pi$-twists) of the $i$th strip. In the present case, they are $L_{00} = 0$
and $L_{11} = + 1$ as drawn in Fig. \ref{bramafold}. The off-diagonal element
$L_{ij}$ ($i \neq j$), corresponds to the permutation between the $i$th and
the $j$th branches. By construction, the matrix is symmetric. The left bracket
corresponds to the splitting chart and the double bracket to the joining 
chart \cite{Ros13}. From this linking matrix and the orbital sequence 
designating periodic orbits, there is an algorithm to compute the linking 
number which quantifies the number of times one orbit circles another one 
(see \cite{LeS94,LetPhD,Let95a} for details). When the linking numbers computed
from the periodic orbits numerically extracted from the attractor are equal to
those algebraically computed from the linker (\ref{linker}), the branched 
manifold is validated and the topological characterization is completed.

\begin{figure}[ht]
  \centering
  \begin{tabular}{ccc}
    \includegraphics[width=0.30\textwidth]{lor2map.eps} & ~~~ & 
    \includegraphics[width=0.40\textwidth]{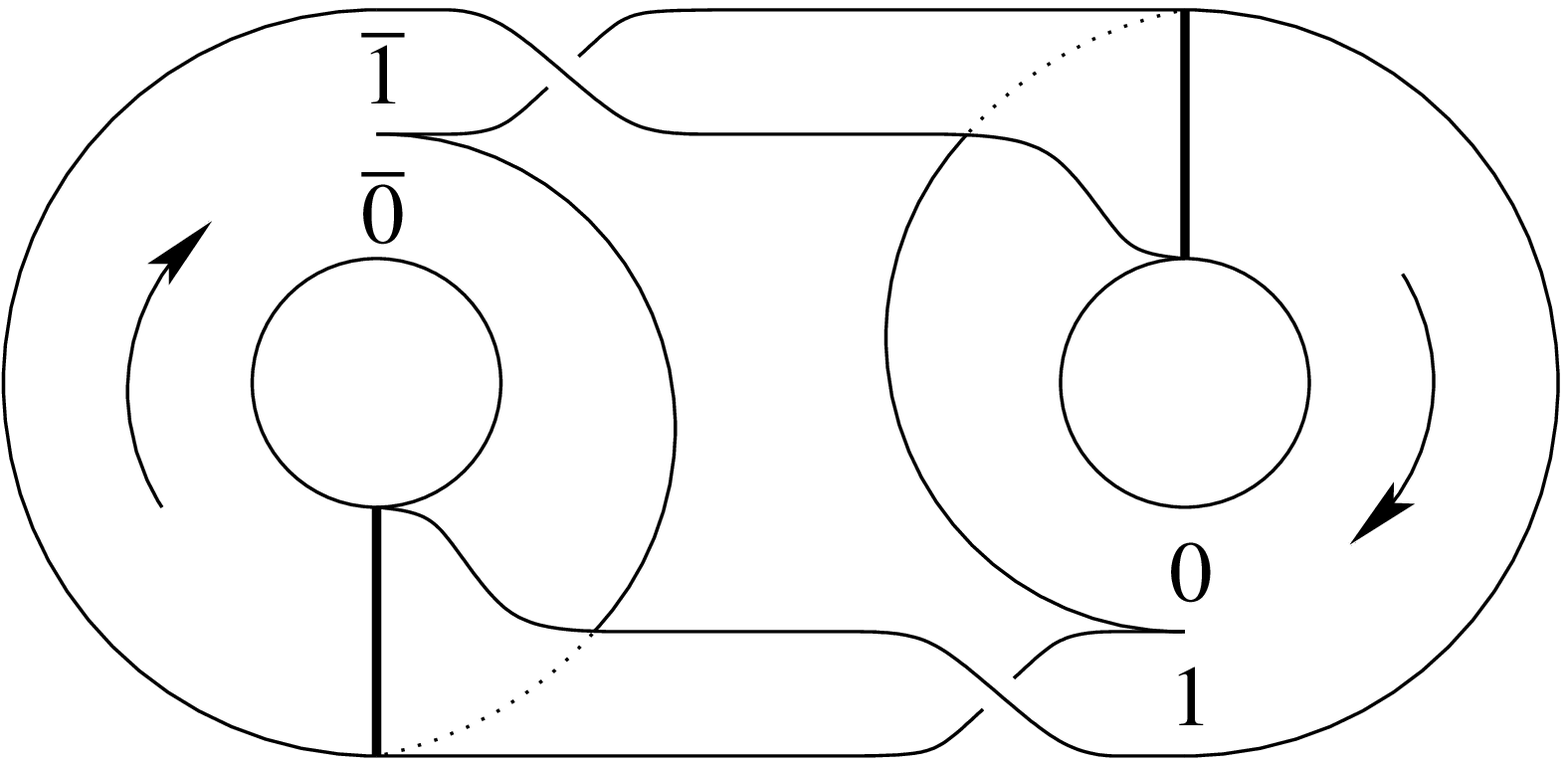} \\[-0.2cm]
  \end{tabular}
  \caption{(a) First-return map to the two-component Poincar\'e section of the 
Lorenz attractor and (b) its description by a branched manifold. Parameter 
values: $R = 28$, $\sigma = 10$, and $b = \frac{8}{3}$.}
  \label{lormap2}
\end{figure}

The Lorenz attractor is bounded by the genus-3 torus drawn in Fig.\ 
\ref{boutos}(b) \cite{Let05c}. The Poincar\'e section is therefore made of the
two-components Poincar\'e section
\begin{equation}
  {\cal P}_\pm \equiv
  \left\{
   (y_n, z_n) \in \mathbb{R}^2 ~|~ 
	x_n = \pm \sqrt{b (R-1)}, \dot{x}_n \lessgtr 0
  \right\} \, . 
\end{equation}
Normalizing these two components in the interval $]-1,0[$ and $]0,+1[$, 
respectively, to construct the variable $\rho$, a four branch first-return map 
is obtained [Fig.\ \ref{lormap2}(a)] \cite{Byr04}. According to this map, the 
branches are ordered according to 
\[ \0 \triangleleft \1 \triangleleft 0 \triangleleft 1 \, . \]
Since the attractor is bounded by a genus-3 torus, there is two joining charts
(one per component of the Poincar\'e section). The branched manifold is drawn 
in Fig.\ \ref{lormap2}(b). It can be described by a double linker
\begin{equation}
  \label{linlor}
  {\cal L}_{\rm L} = 
  \begin{array}{cc}
    \left[
      \begin{array}{cc}
        ~0 & 0 \\
        \cdot & +1 
      \end{array} 
    \right. &
    \left.
      \begin{array}{cc}
        \cdot & \cdot \\
        ~0 & ~0 
      \end{array}
	  \right\rsem \\[0.4cm] 
    \left[
      \begin{array}{cc}
        \cdot & \cdot \\
        ~0 & ~0 
      \end{array} 
    \right. &
    \left.
      \begin{array}{cc}
        ~0 & 0 \\
        \cdot & +1 
      \end{array}
	  \right\rsem 
  \end{array}
\end{equation}
where ``$\cdot$'' means that the $i$th and the $j$th strips do not come from
the same splitting graph and, consequently, cannot be permuted; it also means
that the $i$th strip cannot be reinjected into the $j$th strip as checked with
the Markov transition matrix 
\begin{equation}
  M_{ij} = 
  \left[
    \begin{array}{cccc}
	    0.60 & 0.40 & 0 & 0 \\[0.1cm]
	    0 & 0 & 0.53 & 0.47 \\[0.1cm]
	    0 & 0 & 0.60 & 0.40 \\[0.1cm]
	    0.53 & 0.47 & 0 & 0
    \end{array}
  \right] \, , 
\end{equation}
a feature which is only valid for $i \neq j$ since, for instance, strip 
$\1$ is not reinjected in itself. Notice that we make a difference between
``not permuted'' ($L_{ij} = 0$) and ``cannot be permuted'' ($L_{ij} =$ 
``$\cdot$''): in the former case, strips are in 
the same permutation chart but are not permuted while, in the latter, strips
do not belong to the same permutation chart.

\begin{figure}[ht]
  \centering
  \includegraphics[width=0.35\textwidth]{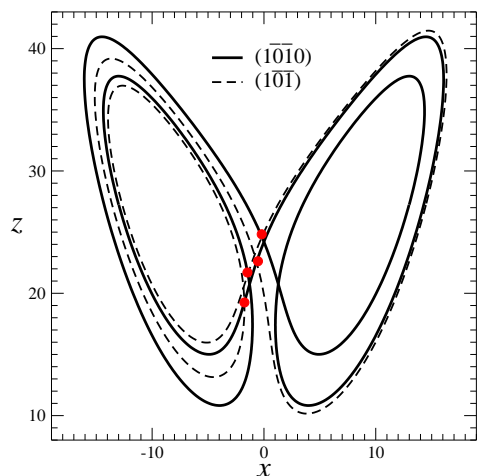} \\[-0.2cm]
	\caption{Periodic orbits $(1\0 \1)$ and $(1 \0 \1 0)$ extracted from 
the Lorenz attractor. Four positive crossings (red disks) are counted between 
these two orbits.}
  \label{lorupos}
\end{figure}

The linker (\ref{linlor}) thus encodes also the transition matrix between the
different strips of the branched manifold: non-removed elements $L_{ij}$ 
indicate that the transition from the $i$th strip to $j$th strip is possible
(even when $L_{ij} = 0 \neq \cdot$). For multi-component attractor, the linker
is thus no longer symmetric. As developed by \cite{LeS94} for one-component
attractor, it is still possible to algebraically compute linking number from a 
multi-component linker. The difference occurs in the joining graph: for a 
($g-1$) component attractor, there is ($g-1$) joining graphs to construct as it 
will be detailed in the example of the two orbits $(1 \0 \1)$ and 
$(1 \0 \1 0)$ (Fig.\ \ref{lorupos}) as follows.

\begin{figure}[ht]
  \centering
  \begin{tabular}{ccc}
	  \includegraphics[height=0.12\textwidth]{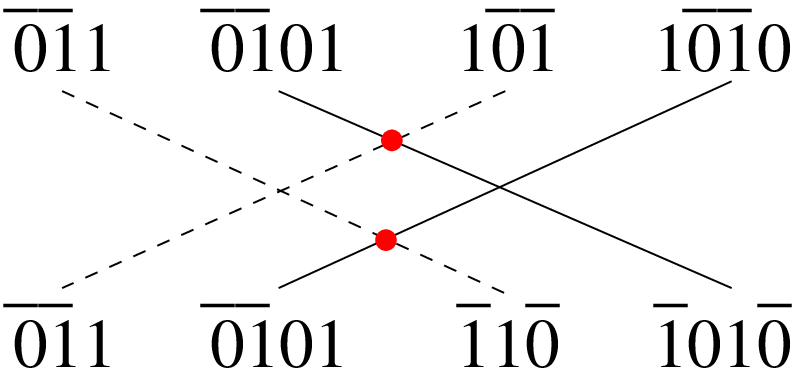} & ~~~ &
    \includegraphics[height=0.12\textwidth]{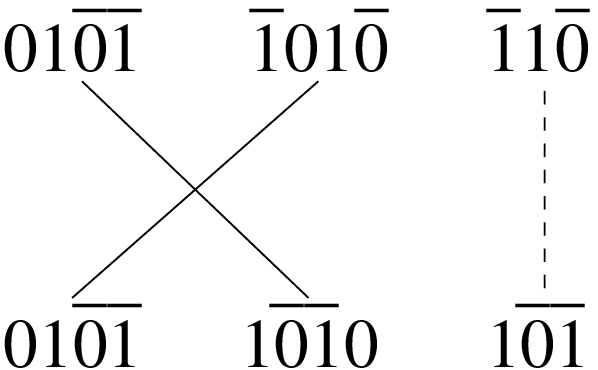} \\
	  (a) $N^1_{\rm join} = +2$ & &
	  (b) $N^2_{\rm join}$ = 0 \\[-0.1cm]
  \end{tabular}
  \caption{The two joining graphs, one for each component of the Poincar\'e
section, between the two orbits $(1 \0 \1)$ and $(1 \0 \1 0)$.}
  \label{joingra}
\end{figure}

Let us start with the numerical extraction of these two orbits. The oriented 
crossings are extracted with the third coordinate
(see \cite{Let95a,Let21d} for details). Four positive crossings are counted:
the linking number between these two periodic orbits is thus
\begin{equation}
	L_{\rm k}(1 \0 \1, 1 \0 \1 0) = + \frac{4}{2} = + 2 \, . 
\end{equation}
Extrapolating the original procedure \cite{LeS94}, the linking number is 
predicted from the linker (\ref{linlor}) by 
\begin{equation}
  L_{\rm k}(1 \0 \1, 1 \0 \1 0) = + \frac{L_{11} + L_{\1 \1} + 2 \, L_{1 \1}
	 + L_{\1 0} + 2 L_{1 \0} + L_{0 \0} + L_{\0 \0} + N^1_{\rm join} 
	 + N^2_{\rm join} }{2} 
\end{equation}
where the three terms $2 L_{\0 \1}$, $L_{0 \0}$ and $L_{\0 \0}$ are removed 
since associated with a ``$\cdot$'' in the linker (\ref{linlor}). The joining 
graphs are constructed
as follows. Let us start with the joining chart between strips $\0$ and $1$. 
This graph is constructed between the periodic points $\0 \1 1$ and $1 \0 \1$
from the period-3 orbit, and from $\0 \1 0 1$ and $1 \0 \1 0$ from the 
period-4 orbit (the first symbol indicates the strip in which the periodic 
point is located). First, these points are ordered according to the kneading 
theory \cite{Bai89,Let95a,LetPhD} as
\[ \0 \1 0 \triangleleft \0 \1 0 1 \triangleleft 1 \0 \1 \triangleleft
1 \0 \1 0 \, . \]
Then, according to \cite{LeS94}, the points $1 \0 \1$ and $1 \0 \1 0$ are 
inverted since belonging to an odd strip. Points from strips $\0$ and $1$ are 
not permuted in block since they come from strips issued from two different 
splitting charts [Fig.\ \ref{lormap2}(b)], that is, from strip which cannot be
permuted. The upper row of the joining chart is thus 
\[ \0 \1 0 \triangleleft \0 \1 0 1 \triangleleft 1 \0 \1 0 \triangleleft
1 \0 \1 \, . \]
The lower row is made of the iterate of the previous points under a Bernoulli
shift. These periodic sequences are then ordered according to the kneading
theory, leading to
\[ \0 \1 0 \triangleleft \0 \1 0 1 \triangleleft \1 1 \0 \triangleleft 
\1 0 1 \0  \, .  \]
It remains to link the sequences related by a Bernoulli shift as shown in 
Fig.\ \ref{joingra}(a) and to count the crossings which are necessarily 
positive \cite{LeS94}. We thus find $N^1_{\rm join} = +2$. Proceeding in a 
similar way with strips $\1$ and $0$, we get $N^2_{\rm join} = 0$
[Fig.\ \ref{joingra}(b)]. The linking number is thus
\begin{equation}
  L_{\rm k} (1 \0 \1, 1 \0 \1 0) 
	= \frac{ + 1 + 1 + 0 + 0 + 0 + 0 + 0 + 2 + 0}{2} 
	= + 2 \, . 
\end{equation}
Counted and predicted linking numbers are thus equal. The here proposed 
procedure for multi-component attractor is thus validated.

When the symmetry of the Lorenz system is removed, the attractor is 
characterized by $m = 1$, $p = 1$, $g = 1$, $n_{\rm c} = 1$ and a branched
manifold as shown in Fig.\ \ref{bramafold} (see \cite{Let21d} for details).
The single difference is that the folding of the R\"ossler attractor is 
replaced with a tearing \cite{Byr04}. Since the bifurcation diagram induced
by a tearing is very different from the one obtained with a folding (a Logistic map) \cite{Gil15}, the have to be distinguished as two different taxa. As 
developed by \cite{Gil07}, different taxa can be easily developed from a known
one but with various symmetries.

\section{Conclusion}
\label{conc}

After more than a half century of scientific researches with computers, a 
plethora of chaotic attractors are today found in the literature. In spite of 
this, what make them specific was not yet very clear and, commonly, their 
characterization is limited to the spectrum of Lyapunov exponents, the first 
two taxonomic ranks according to our views. With the taxonomy here developed, 
we provided some guidelines to complete a full 
characterization of chaotic attractors. We incorporated bounding tori as the 
third taxonomic rank: if this is quite easy for more three-dimensional chaos,
determining bounding tori for higher-dimenisonal dynamics appears as an open
problem in many cases. We left for future works the promising results obtained 
with the Betti numbers and their extension \cite{Cha20,Cha21}.

Even worse is the problem related to the modality defined by the number of 
mixing processes in action in a given dynamics. If, for strongly dissipative 
three-dimensional systems, this is simply quantified with the number of 
monotone branches in first-return map, this is a complex problem 
for weakly dissipative systems and nearly never addressed for conservative 
systems. This is in fact deeply connected to the question of finding a 
generating partition for conservative dynamics \cite{Chr95}. The last taxonomic
rank is associated with the topological description of the structure underlying
the invariant set. We left untouched this last rank, mostly because this is 
still an open problem for most of toroidal chaos, in spite of the recent 
success for two simple cases \cite{Man21a}. 

In spite of all of these problems, the taxonomy we here propose provides 
already a systematic labeling for every type of dynamical regimes based on
two integers, intimately related to the number of positive and null Lyapunov
exponents, respectively. This is a first efficient classification of chaotic
regimes. Determining if the dynamics is degenerated and/or conjugated is 
also an important additional clue, showing that the Lyapunov exponents are 
not always sufficient to identify correctly the type of chaos. Clearly the 
role of bounding tori is fundamental in such a taxonomy and deserves to be 
better promoted. Hopefully, this taxonomy will stimulate dynamicists to go 
deeper in their analysis of the observed behaviors. One of the important 
lessons brought by this study is to show that characterizing chaotic attractors
is still in its infancy. This taxonomy provides some guidelines for developing 
this area of research.

\nonumsection{Acknowledgments} \noindent
C. L. wishes to thank Jean-Marc Malasoma for stimulating discussions.
N. S. thanks Igor Sataev and Vyacheslav Kruglov for fruitful discussion and the
Russian Foundation for Basic Research for financial support of her work 
(grant No. 19-31-60030).

\bibliographystyle{ws-ijbc}
\bibliography{SysDyn}

\begin{thebibliography}{59}
\newcommand{\enquote}[1]{``#1''}
\providecommand{\natexlab}[1]{#1}
\providecommand{\url}[1]{\texttt{#1}}
\providecommand{\urlprefix}{URL }
\expandafter\ifx\csname urlstyle\endcsname\relax
  \providecommand{\doi}[1]{doi:\discretionary{}{}{}#1}\else
  \providecommand{\doi}{doi:\discretionary{}{}{}\begingroup
  \urlstyle{rm}\Url}\fi

\bibitem[{Anishchenko \& Nikolaev(2005)}]{Ani05}
Anishchenko, V.~S. \& Nikolaev, S.~M. [2005] \enquote{Generator of
  quasi-periodic oscillations featuring two-dimensional torus doubling
  bifurcations,} \emph{Technical Physics Letters} \textbf{31},  853--855,
  \doi{10.1134/1.2121837}.

\bibitem[{Arn\'eodo \emph{et~al.}(1983)Arn\'eodo, Coullet \& Spiegel}]{Arn83}
Arn\'eodo, A., Coullet, P.~H. \& Spiegel, E.~A. [1983] \enquote{Cascade of
  period doublings of tori,} \emph{Physics Letters A} \textbf{94},  1 -- 6,
  \doi{10.1016/0375-9601(83)90272-4}.

\bibitem[{Aziz-Alaoui(1999)}]{Azi99}
Aziz-Alaoui, M.~A. [1999] \enquote{Differential equations with multispiral
  attractors,} \emph{International Journal of Bifurcation $\&$ Chaos}
  \textbf{9},  1009--1039, \doi{10.1142/S0218127499000729}.

\bibitem[{Bai-Lin(1989)}]{Bai89}
Bai-Lin, H. [1989] \emph{Elementary symbolic dynamics and chaos in dissipative
  systems} (World Scientific Publishing, Singapore).

\bibitem[{Birkhoff(1927)}]{Bir27}
Birkhoff, G.~D. [1927] \emph{Dynamical Systems} (American Mathematical Society,
  New York).

\bibitem[{Byrne \emph{et~al.}(2004)Byrne, Gilmore \& Letellier}]{Byr04}
Byrne, G., Gilmore, R. \& Letellier, C. [2004] \enquote{Distinguishing between
  folding and tearing mechanisms in strange attractors,} \emph{Physical Review
  E} \textbf{70},  056214, \doi{10.1103/PhysRevE.70.056214}.

\bibitem[{Char{\'o} \emph{et~al.}(2020)Char{\'o}, Artana \&
  Sciamarella}]{Cha20}
Char{\'o}, G.~D., Artana, G. \& Sciamarella, D. [2020] \enquote{Topology of
  dynamical reconstructions from {L}agrangian data,} \emph{Physica D}
  \textbf{405},  132371, \doi{10.1016/j.physd.2020.132371}.

\bibitem[{Char{\'o} \emph{et~al.}(2021)Char{\'o}, Artana \&
  Sciamarella}]{Cha21}
Char{\'o}, G.~D., Artana, G. \& Sciamarella, D. [2021] \enquote{Topological
  colouring of fluid particles unravels finite-time coherent sets,}
  \emph{Journal of Fluid Mechanics} \textbf{923},  A17,
  \doi{10.1017/jfm.2021.561}.

\bibitem[{Char\'o \emph{et~al.}(2019)Char\'o, Sciamarella, Mangiarotti, Artana
  \& Letellier}]{Cha19}
Char\'o, G.~D., Sciamarella, D., Mangiarotti, S., Artana, G. \& Letellier, C.
  [2019] \enquote{Equivalence between the unsteady double-gyre system and a
  {4D} autonomous conservative chaotic system,} \emph{Chaos} \textbf{29},
  123126, \doi{10.1063/1.5120625}.

\bibitem[{Christiansen \& Politi(1995)}]{Chr95}
Christiansen, F. \& Politi, A. [1995] \enquote{Generating partition for the
  standard map,} \emph{Physical Review E} \textbf{51},  R3811--R3814,
  \doi{10.1103/PhysRevE.51.R3811}.

\bibitem[{Curry \& Yorke(1978)}]{Cur78}
Curry, J. \& Yorke, J.~A. [1978] \enquote{A transition from {H}opf bifurcation
  to chaos: computer experiments with maps on $\mathbb{R}^2$,} \emph{Lecture
  Notes in Mathematics} \textbf{668},  48--66.

\bibitem[{Ereshefsky(2000)}]{Ere00}
Ereshefsky, M. [2000] \emph{The Poverty of the {L}innaean Hierarchy: A
  Philosophical Study of Biological Taxonomy}, Cambridge Studies in Philosophy
  and Biology (Cambridge University Press, Cambdrige),
  \doi{10.1017/CBO9780511498459}.

\bibitem[{Franceschini(1983)}]{Fra83}
Franceschini, V. [1983] \enquote{Bifurcations of tori and phase locking in a
  dissipative system of differential equations,} \emph{Physica D} \textbf{6},
  285--304, \doi{10.1016/0167-2789(83)90013-1}.

\bibitem[{Frederickson \emph{et~al.}(1983)Frederickson, Kaplan, Yorke \&
  Yorke}]{Fre83}
Frederickson, P., Kaplan, J.~L., Yorke, E.~D. \& Yorke, J.~A. [1983]
  \enquote{The {L}yapunov dimension of strange attractors,} \emph{Journal of
  Differential Equations} \textbf{49},  185--207,
  \doi{10.1016/0022-0396(83)90011-6}.

\bibitem[{Gilmore(1998)}]{Gil98}
Gilmore, R. [1998] \enquote{Topological analysis of chaotic dynamical systems,}
  \emph{Reviews of Modern Physics} \textbf{70},  1455--1529,
  \doi{10.1103/RevModPhys.70.1455}.

\bibitem[{Gilmore(2015)}]{Gil15}
Gilmore, R. [2015] \enquote{Explosions in {L}orenz maps,} \emph{Chaos, Solitons
  $\&$ Fractals} \textbf{76},  130--140.

\bibitem[{Gilmore \& Lefranc(2003)}]{Gil03}
Gilmore, R. \& Lefranc, M. [2003] \emph{The topology of chaos} (Wiley), ISBN
  9780471408161, \doi{10.1002/9783527617319}.

\bibitem[{Gilmore \& Letellier(2007)}]{Gil07}
Gilmore, R. \& Letellier, C. [2007] \emph{The symmetry of chaos} (Oxford
  University Press, New York).

\bibitem[{Grebogi \emph{et~al.}(1984)Grebogi, Ott, Pelikan \& Yorke}]{Gre84}
Grebogi, C., Ott, E., Pelikan, S. \& Yorke, J.~A. [1984] \enquote{Strange
  attractors that are not chaotic,} \emph{Physica D} \textbf{13},  261--268.

\bibitem[{H\'enon \& Heiles(1964)}]{Hen64}
H\'enon, M. \& Heiles, C. [1964] \enquote{The applicability of the third
  integral of motion: some numerical experiments,} \emph{The astronomical
  Journal} \textbf{69},  73--79.

\bibitem[{Kaneko(1983)}]{Kan83}
Kaneko, K. [1983] \enquote{{Doubling of torus},} \emph{Progress of Theoretical
  Physics} \textbf{69},  1806--1810, \doi{10.1143/PTP.69.1806}.

\bibitem[{Kaneko(1986)}]{Kan86}
Kaneko, K. [1986] \emph{Collapse of tori and genesis of chaos in dissipative
  systems} (World Scientific Publishing, Singapore).

\bibitem[{Kaplan \& Yorke(1979)}]{Kap79}
Kaplan, J. \& Yorke, J. [1979] \enquote{Chaotic behavior of multidimensional
  difference equations,} \emph{Lecture Notes in Mathematics} \textbf{730},
  204--227.

\bibitem[{Klein(1882)}]{Kle82}
Klein, F. [1882] \emph{\"Uber Riemann's Theorie der algebraischen Functionen
  und ihrer Integrale} (Druck und Verlag von B. G. Teubner, Leipzig).

\bibitem[{Klein \& Baier(1991)}]{Kle91}
Klein, M. \& Baier, G. [1991] \enquote{Hierarchies of dynamical systems,}
  \emph{A chaotic hierarchy} (World Scientific Publishing), pp. 1--24.

\bibitem[{Le~Sceller \emph{et~al.}(1994)Le~Sceller, Letellier \&
  Gouesbet}]{LeS94}
Le~Sceller, L., Letellier, C. \& Gouesbet, G. [1994] \enquote{Algebraic
  evaluation of linking numbers of unstable periodic orbits in chaotic
  attractors,} \emph{Physical Review E} \textbf{49},  4693--4695,
  \doi{10.1103/PhysRevE.49.4693}.

\bibitem[{Letellier(1994)}]{LetPhD}
Letellier, C. [1994] \enquote{Caract\'erisation topologique et reconstruction
  des attracteurs \'etranges,}  PhD thesis, University of Paris VII, Paris,
  France, \doi{10.13140/RG.2.1.1280.5281}.

\bibitem[{Letellier(2021)}]{Let21d}
Letellier, C. [2021] \enquote{Branched manifolds for the three types of
  unimodal maps,} \emph{Communications in Nonlinear Science and Numerical
  Simulation} \textbf{101},  105869, \doi{10.1016/j.cnsns.2021.105869}.

\bibitem[{Letellier \& Aguirre(2012)}]{Let12}
Letellier, C. \& Aguirre, L.~A. [2012] \enquote{Required criteria for
  recognizing new types of chaos: Application to the ``cord'' attractor,}
  \emph{Physical Review E} \textbf{85},  036204.

\bibitem[{Letellier \emph{et~al.}(2007)Letellier, Bennoud \& Martel}]{Let07a}
Letellier, C., Bennoud, M. \& Martel, G. [2007] \enquote{Intermittency and
  period-doubling cascade on tori in a bimode laser model,} \emph{Chaos,
  Solitons $\&$ Fractals} \textbf{33},  782--794,
  \doi{10.1016/j.chaos.2006.01.109}.

\bibitem[{Letellier \emph{et~al.}(1994)Letellier, Dutertre \&
  Gouesbet}]{Let94a}
Letellier, C., Dutertre, P. \& Gouesbet, G. [1994] \enquote{Characterization of
  the {L}orenz system, taking into account the equivariance of the vector
  field,} \emph{Physical Review E} \textbf{49},  3492--3495,
  \doi{10.1103/PhysRevE.49.3492}.

\bibitem[{Letellier \emph{et~al.}(1995)Letellier, Dutertre \& Maheu}]{Let95a}
Letellier, C., Dutertre, P. \& Maheu, B. [1995] \enquote{Unstable periodic
  orbits and templates of the {R}\"ossler system: {T}oward a systematic
  topological characterization,} \emph{Chaos} \textbf{5},  271--282,
  \doi{10.1063/1.166076}.

\bibitem[{Letellier \& Gilmore(2001)}]{Let01}
Letellier, C. \& Gilmore, R. [2001] \enquote{Covering dynamical systems:
  Two-fold covers,} \emph{Physical Review E} \textbf{63},  016206,
  \doi{10.1103/PhysRevE.63.016206}.

\bibitem[{Letellier \& Gilmore(2009)}]{Let09c}
Letellier, C. \& Gilmore, R. [2009] \enquote{Poincar\'e sections for a new
  three-dimensional toroidal attractor,} \emph{Journal of Physics A}
  \textbf{42},  015101.

\bibitem[{Letellier \& Gouesbet(1995)}]{Let95d}
Letellier, C. \& Gouesbet, G. [1995] \enquote{Topological characterization of a
  system with high-order symmetries: the proto-{L}orenz system,} \emph{Physical
  Review E} \textbf{52},  4754--4761.

\bibitem[{Letellier \& R\"ossler(2020)}]{Let20b}
Letellier, C. \& R\"ossler, O.~E. [2020] ``An updated hierarchy of chaos,''
  \emph{Chaos: The world of nonperiodic oscillations} (Springer, Cham,
  Switzerland), pp. 181--203, \doi{10.1007/978-3-030-44305-4}.

\bibitem[{Letellier \emph{et~al.}(2005)Letellier, Tsankov, Byrne \&
  Gilmore}]{Let05c}
Letellier, C., Tsankov, T.~D., Byrne, G. \& Gilmore, R. [2005]
  \enquote{Large-scale structural reorganization of strange attractors,}
  \emph{Physical Review E} \textbf{72},  026212,
  \doi{10.1103/PhysRevE.72.026212}.

\bibitem[{Malykh \emph{et~al.}(2020)Malykh, Bakhanova, Kazakov, Pusuluri \&
  Shilnikov}]{Maly20}
Malykh, S., Bakhanova, Y., Kazakov, A., Pusuluri, K. \& Shilnikov, A. [2020]
  \enquote{Homoclinic chaos in the {R\"o}ssler model,} \emph{Chaos}
  \textbf{30},  113126, \doi{10.1063/5.0026188}.

\bibitem[{Mangiarotti \& Letellier(2021)}]{Man21a}
Mangiarotti, S. \& Letellier, C. [2021] \enquote{Topological characterization
  of toroidal chaos: A branched manifold for the {De}ng toroidal attractor,}
  \emph{Chaos} \textbf{31},  013129, \doi{10.1063/5.0025924}.

\bibitem[{M{\'e}nard \emph{et~al.}(2000)M{\'e}nard, Letellier, Maquet, Sceller
  \& Gouesbet}]{Men00}
M{\'e}nard, O., Letellier, C., Maquet, J., Sceller, L.~L. \& Gouesbet, G.
  [2000] \enquote{Analysis of a non synchronized sinusoidally driven dynamical
  system,} \emph{International Journal of Bifurcation $\&$ Chaos} \textbf{10},
  1759--1772.

\bibitem[{Mindlin \& Gilmore(1992)}]{Min92}
Mindlin, G.~M. \& Gilmore, R. [1992] \enquote{Topological analysis and
  synthesis of chaotic time series,} \emph{Physica D} \textbf{58},  229--242,
  \doi{10.1016/0167-2789(92)90111-Y}.

\bibitem[{Miranda \& Stone(1993)}]{Mir93}
Miranda, R. \& Stone, E. [1993] \enquote{The proto-{L}orenz system,}
  \emph{Physics Letters A} \textbf{178},  105--113.

\bibitem[{Parker \& Chua(1989)}]{Par89}
Parker, T.~S. \& Chua, L.~O. [1989] \emph{Practical numerical algorithms for
  chaotic systems} (Springer Verlag, Berlin--New York).

\bibitem[{Rosalie \& Letellier(2013)}]{Ros13}
Rosalie, M. \& Letellier, C. [2013] \enquote{Systematic template extraction
  from chaotic attractors: {\sc i}. {G}enus-one attractors with an inversion
  symmetry,} \emph{Journal of Physics A} \textbf{46},  375101,
  \doi{10.1088/1751-8113/46/37/375101}.

\bibitem[{R{\"o}ssler(1976)}]{Ros76c}
R{\"o}ssler, O.~E. [1976] \enquote{An equation for continuous chaos,}
  \emph{Physics Letters A} \textbf{57},  397--398,
  \doi{10.1016/0375-9601(76)90101-8}.

\bibitem[{R{\"o}ssler(1978)}]{Ros78b}
R{\"o}ssler, O.~E. [1978] \enquote{The hierarchy of chaos,}  \emph{10th Summer
  Seminar on Applied Mathematics}, {F. C. Ho}ppensteadt (organizer), University
  of Utah.

\bibitem[{R\"ossler(1979{\natexlab{a}})}]{Ros79b}
R\"ossler, O.~E. [1979{\natexlab{a}}] \enquote{Continuous chaos: four prototype
  equations,} \emph{Annals of the New York Academy of Sciences} \textbf{316},
  376--392.

\bibitem[{R\"ossler(1979{\natexlab{b}})}]{Ros79a}
R\"ossler, O.~E. [1979{\natexlab{b}}] \enquote{An equation for hyperchaos,}
  \emph{Physics Letters A} \textbf{71},  155--157.

\bibitem[{R\"ossler(1983)}]{Ros83}
R\"ossler, O.~E. [1983] \enquote{The chaotic hierarchy,} \emph{Zeitschrift
  f\"ur Naturforschung A} \textbf{38},  788--801.

\bibitem[{Ruelle \& Takens(1971)}]{Rue71}
Ruelle, D. \& Takens, F. [1971] \enquote{On the nature of turbulence,}
  \emph{Communications in Mathematical Physics} \textbf{20},  167--192.

\bibitem[{S\'equin(2013)}]{Seq13}
S\'equin, C.~H. [2013] \enquote{On the number of {Kl}ein bottle types,}
  \emph{Journal of Mathematics and the Arts} \textbf{7},  51--63,
  \doi{10.1080/17513472.2013.795883}.

\bibitem[{Sprott(1997)}]{Spr97b}
Sprott, J.~C. [1997] \enquote{Some simple chaotic jerk functions,}
  \emph{American Journal of Physics} \textbf{65},  537--543,
  \doi{10.1119/1.18585}.

\bibitem[{Sprott(2011)}]{Spr11}
Sprott, J.~C. [2011] \enquote{A proposed standard for the publication of new
  chaotic systems,} \emph{International Journal of Bifurcation $\&$ Chaos}
  \textbf{21},  2391--2394, \doi{10.1142/S021812741103009X}.

\bibitem[{Stankevich \emph{et~al.}(2020)Stankevich, Shchegoleva, Sataev \&
  Kuznetsov}]{Sta20b}
Stankevich, N.~V., Shchegoleva, N.~A., Sataev, I.~R. \& Kuznetsov, A.~P. [2020]
  \enquote{Three-dimensional torus breakdown and chaos with two zero {Ly}apunov
  exponents in coupled radio-physical generators,} \emph{Journal of
  Computational and Nonlinear Dynamics} \textbf{15},  111001.

\bibitem[{Tricoche \emph{et~al.}(2011)Tricoche, Garth \& Sanderson}]{Tri11}
Tricoche, X., Garth, C. \& Sanderson, A. [2011] \enquote{Visualization of
  topological structures in area-preserving maps,} \emph{IEEE Transactions on
  Visualization and Computer Graphics} \textbf{17},  1765--1774,
  \doi{10.1109/TVCG.2011.254}.

\bibitem[{Tsankov \& Gilmore(2003)}]{Tsa03}
Tsankov, T.~D. \& Gilmore, R. [2003] \enquote{Strange attractors are classified
  by bounding tori,} \emph{Physical Review Letters} \textbf{91},  134104.

\bibitem[{Tsankov \& Gilmore(2004)}]{Tsa04}
Tsankov, T.~D. \& Gilmore, R. [2004] \enquote{Topological aspects of the
  structure of chaotic attractors in $\mathbb{R}^3$,} \emph{Physical Review E}
  \textbf{69},  056206.

\bibitem[{Tufillaro \emph{et~al.}(1992)Tufillaro, Abbott \& Reilly}]{Tuf92}
Tufillaro, N.~B., Abbott, T. \& Reilly, J. [1992] \emph{An experimental
  approach to nonlinear dynamics and chaos} (Addison-Wesley, Redwood City, CA),
  ISBN 9780201554410.

\bibitem[{Ueda(1993)}]{Ued93}
Ueda, Y. [1993] \emph{The Road to Chaos}, Science frontier express series
  (Aerial Press, Santa Cruz, CA), ISBN 9780942344141.

\end{thebibliography}

\end{document}